\documentclass[a4paper,11pt]{article}
\pdfoutput=1
\usepackage{jheppub}

\usepackage[utf8]{inputenc}
\usepackage[T1]{fontenc}
\usepackage{lmodern}
\usepackage{textcomp}
\usepackage{microtype}

\usepackage[font=small]{caption}
\usepackage[font=footnotesize]{subcaption}
\newcommand*\sublabel[1]{\begin{subfigure}{0pt}\captionsetup{labelformat=empty,skip=0pt}\caption{}\label{#1}\end{subfigure}}

\graphicspath{{figs/}}

\usepackage{booktabs}
\usepackage{multirow}

\usepackage{units}
\newcommand{\order}[1]{\mathcal O\left(#1\right)}

\makeatletter
\g@addto@macro\bfseries{\boldmath}
\makeatother

\usepackage{xspace}
\newcommand{\software}[2][]{\texttt{#2}\xspace#1}
\newcommand*{\doi}[1]{doi: \href{http://dx.doi.org/#1}{#1}}

\usepackage{array}
\newcommand{\undefcolumntype}[1]{\expandafter\let\csname NC@find@#1\endcsname\relax}
\newcommand{\forcenewcolumntype}[1]{\undefcolumntype{#1}\newcolumntype{#1}}
\AtBeginDocument{\forcenewcolumntype{|}{@{\hskip\tabcolsep}}}

\title{Heavy Higgs Bosons at Low $\tan\beta$:\\from the LHC to \unit[100]{TeV}}
\author[a]{Nathaniel Craig,}
\author[b,c]{Jan Hajer,}
\author[b]{Ying-Ying Li,}
\author[b]{Tao Liu}
\author[a]{and Hao Zhang}

\emailAdd{ncraig@physics.ucsb.edu}
\emailAdd{jan.hajer@ust.hk}
\emailAdd{ylict@connect.ust.hk}
\emailAdd{taoliu@ust.hk}
\emailAdd{zhanghao@physics.ucsb.edu}

\affiliation[a]{Department of Physics, University of California, Santa Barbara, CA 93106, USA}
\affiliation[b]{Department of Physics, The Hong Kong University of Science and Technology,
\\ Clear Water Bay, Kowloon, Hong Kong S.A.R., P.R.C.}
\affiliation[c]{Institute for Advanced Study, The Hong Kong University of Science and Technology,
\\ Clear Water Bay, Kowloon, Hong Kong S.A.R., P.R.C.}

\abstract{%
We present strategies to search for heavy neutral Higgs bosons decaying to top quark pairs, as often occurs at low $\tan\beta$ in type~II two Higgs doublet models such as the Higgs sector of the MSSM.
The resonant production channel is unsatisfactory due to interference with the SM background.
We instead propose to utilize same-sign dilepton signatures arising from the production of heavy Higgs bosons in association with one or two top quarks and subsequent decay to a top pair.
We find that for heavier neutral Higgs bosons the production in association with one top quark provides greater sensitivity than production in association with two top quarks.
We obtain current limits at the LHC using Run I data at \unit[8]{TeV} and forecast the sensitivity of a dedicated analysis during Run II at \unit[14]{TeV}.
Then we perform a detailed BDT study for the \unit[14]{TeV} LHC and a future \unit[100]{TeV} collider.
\hfill\url{https://github.com/BoostedColliderAnalysis/BoCA}
\bigskip \\ \today
}

\begin{document}

\maketitle

\section{Introduction} \label{sec:int}

The search for additional Higgs bosons is a high priority for current and future colliders.
The ATLAS and CMS collaborations have performed searches for heavy neutral~\cite{Aad:2012cfr, Aad:2014vgg, Aad:2015wra, Aad:2015kna, Aad:2015tna, Aad:2015agg, Khachatryan:2014jya, Khachatryan:2015tra, CMS:2015ooa}
and charged~\cite{Aad:2014kga, Aad:2015nfa, Khachatryan:2015qxa} Higgs bosons during the first run of the Large Hadron Collider (LHC) at $\sqrt{s} = \unit[8]{TeV}$.
The reach of such searches will expand considerably in the future; in its second run, the LHC has started to take data at $\sqrt{s} = \unit[13]{TeV}$, with an eventual goal of collisions at $\sqrt{s} = \unit[14]{TeV}$.
In the long run, the center-of-mass energy of next-generation $pp$-colliders can be as much as an order of magnitude higher than that of the LHC. Such a high energy scale opens up new signal channels which are suppressed at lower energies.

Among the most challenging scenarios are those in which heavy neutral Higgs bosons decay predominantly into $t \bar t$ pairs.
Dedicated studies probing the heavy Higgs sector via $t \bar t$ final states have recently been presented for
the LHC~\cite{Dev:2014yca, Craig:2015jba, Hajer:2015gka, Bhattacherjee:2015sga, Gori:2016zto} and a \unit[100]{TeV} collider~\cite{Hajer:2015gka}.
In motivated models for extended Higgs sectors described by a type~II two Higgs doublet model (2HDM), the moderate $\tan\beta$ region can be covered up to \unit[1]{TeV} using BDT methods, in large part by using bottom quark associated production.
However, the low $\tan\beta$ region remains thus far uncovered as the conventional resonant $t\bar t$ channel suffers from interference with the Standard Model background~\cite{Gaemers:1984sj, Dicus:1994bm, Frederix:2007gi, Jung:2015gta} and the cross section for bottom quark associated production is negligible. This motivates dedicated searches for signals in the low $\tan \beta$ region; the development of strategies for such searches is the goal of this paper.

\begin{figure}
\begin{subfigure}{.5\textwidth}
\includegraphics[width=\textwidth]{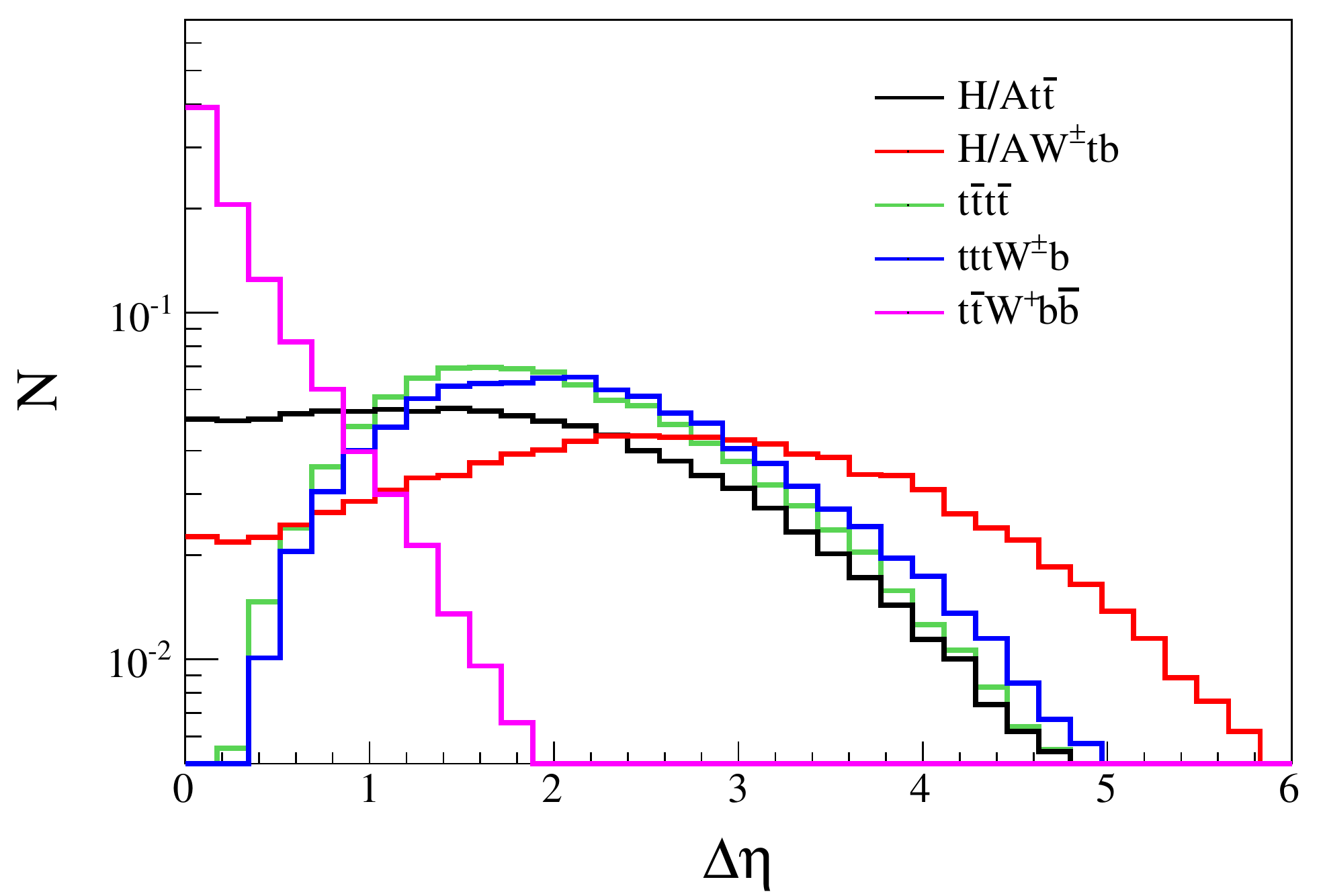}
\caption{\unit[500]{GeV} heavy Higgs at \unit[14]{TeV}}
\label{fig:delta eta lhc}
\end{subfigure}
\begin{subfigure}{.5\textwidth}
\includegraphics[width=\textwidth]{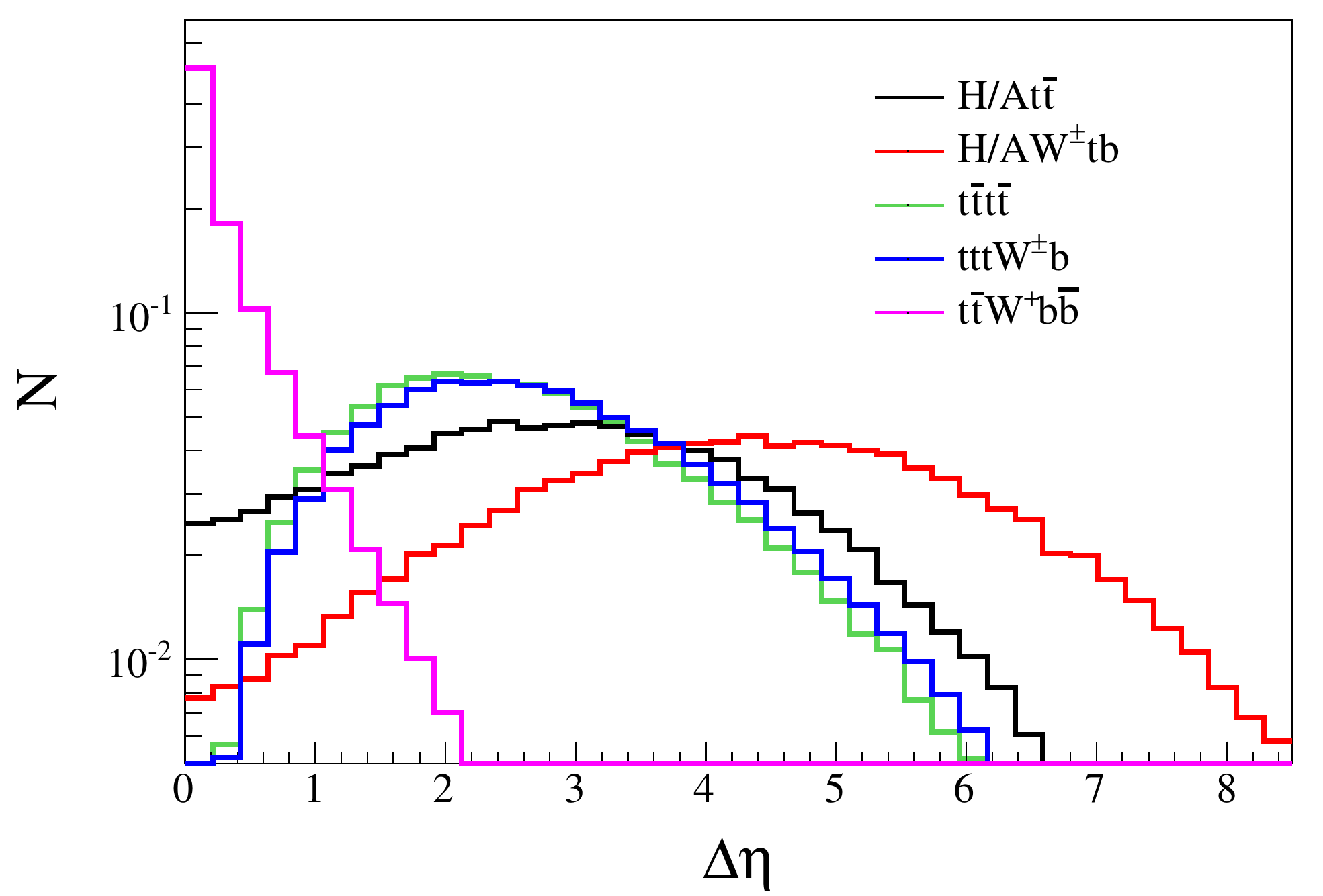}
\caption{\unit[5]{TeV} heavy Higgs at \unit[100]{TeV}}
\label{fig:deltaEta 100}
\end{subfigure}
\caption{%
(\subref{fig:delta eta lhc})  Distribution of the rapidity difference between two bottom quarks
at \unit[14]{TeV} for a \unit[500]{GeV} heavy Higgs. The rapidity difference is calculated between the bottom pair found in the decay products of the particles produced
in association with the heavy Higgs, which involves either one (red) or two (black) top quarks.
(\subref{fig:deltaEta 100}) Distribution of the rapidity difference  at \unit[100]{TeV} for a \unit[5]{TeV} heavy Higgs. For comparison, in each case we also show the maximal rapidity difference between all $b$-quarks in the $t\bar tt\bar t$ background (green),
as well as between the soft additional $b$ and one of the $b$-quarks coming from a top decay in the $tt tW^\pm b$ background (blue).
Additionally, we show the rapidity difference between the two $b$ quarks in the $t\bar tW^\pm b\bar b$ background (pink).
}
\label{fig:delta eta}
\end{figure}

\begin{figure}
\begin{subfigure}{.5\textwidth}
\includegraphics[width=\textwidth]{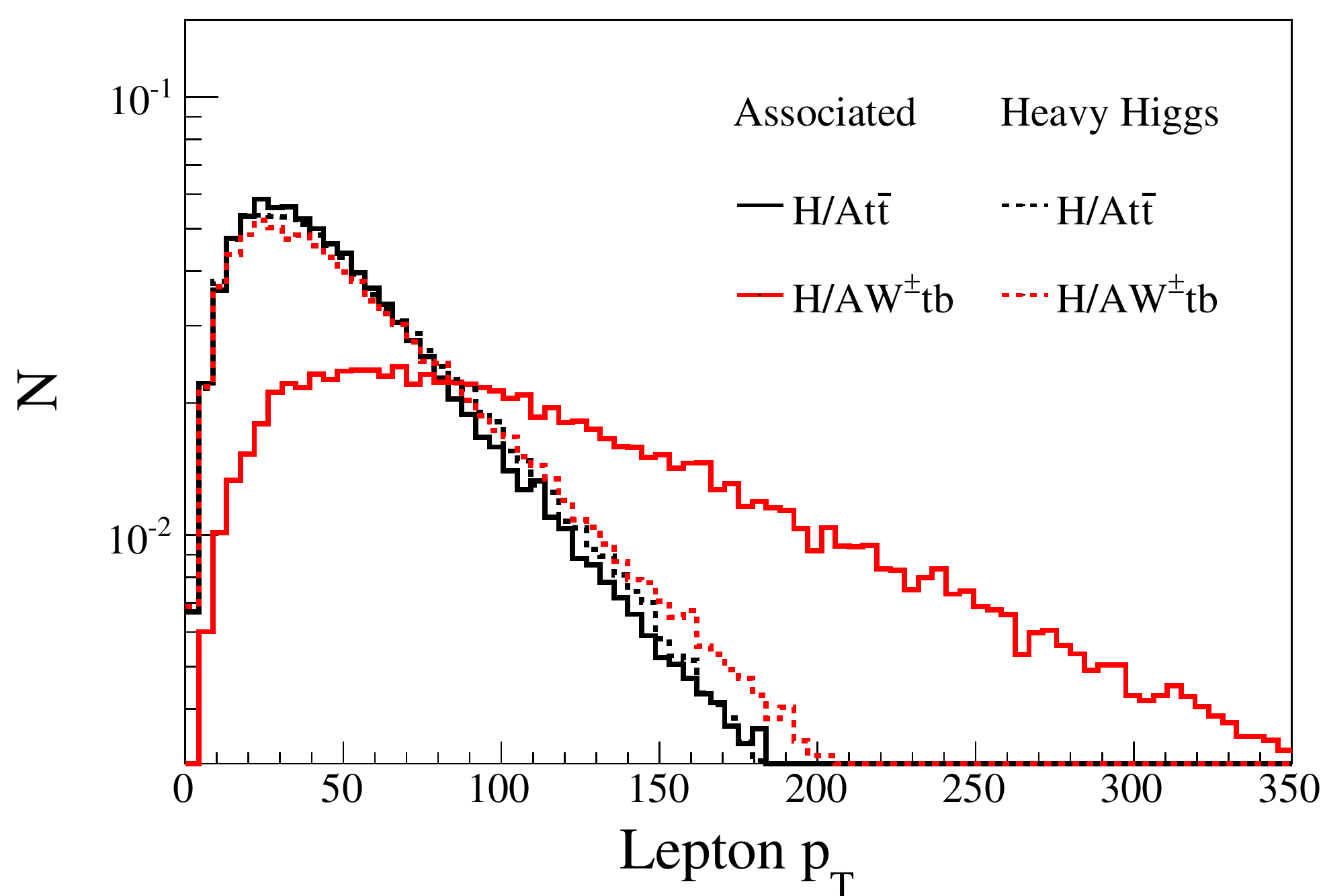}
\caption{\unit[500]{GeV} heavy Higgs at \unit[14]{TeV}}
\label{fig:lepton pt lhc}
\end{subfigure}
\begin{subfigure}{.5\textwidth}
\includegraphics[width=\textwidth]{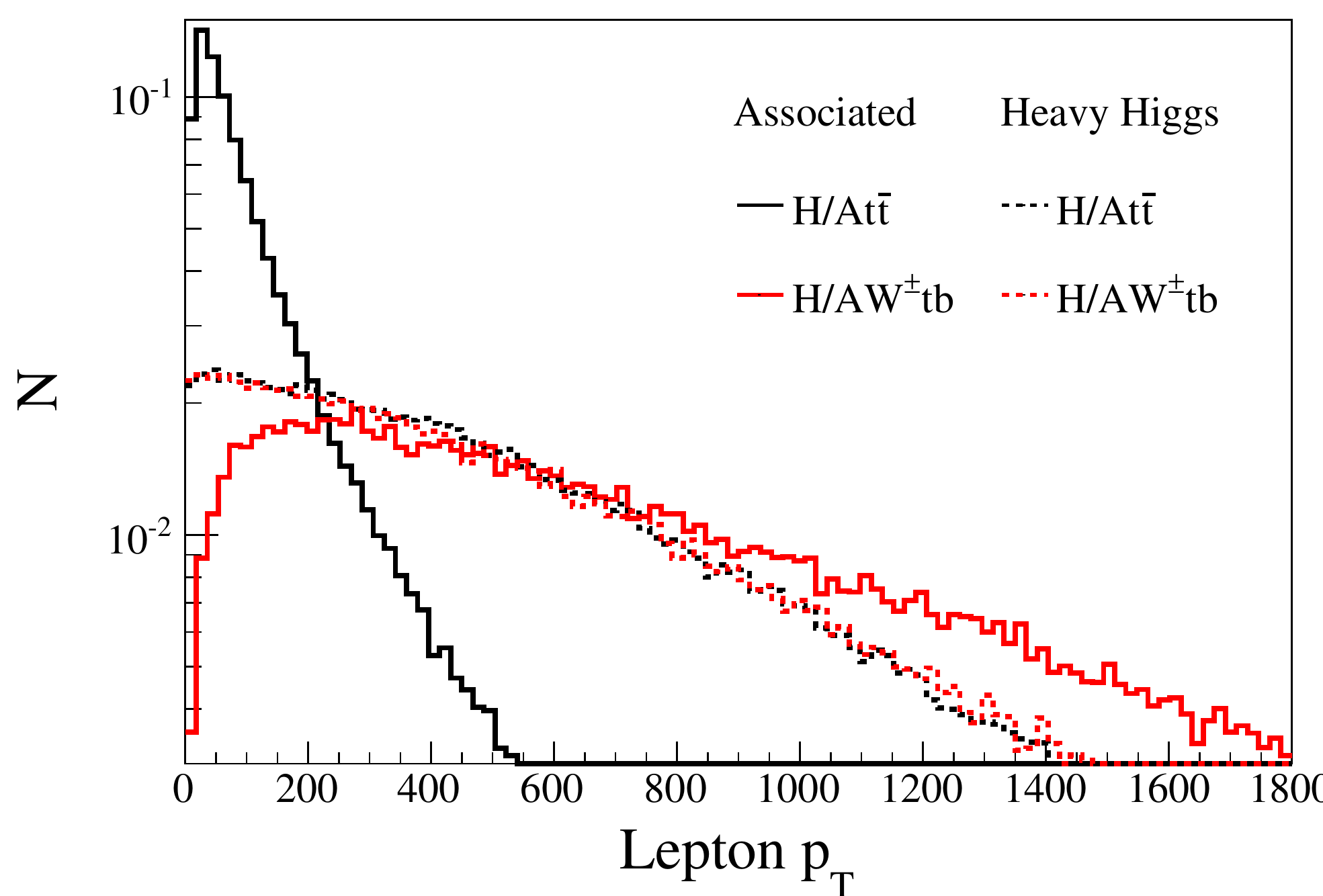}
\caption{\unit[5]{TeV} heavy Higgs at \unit[100]{TeV}}
\label{fig:lepton pt 100}
\end{subfigure}
\caption{%
(\subref{fig:lepton pt lhc}) Transverse momenta of the leptons at \unit[14]{TeV} for a \unit[500]{GeV} heavy Higgs. The transverse momenta are plotted for the two signal channels $H/At\bar t$ (black) and $H/AW^\pm tb$ (red) for cases in which the lepton comes from the decay products of a heavy Higgs boson (dotted) or the decay products of an associated particle ($W$ boson for three-top case) (solid).
(\subref{fig:lepton pt 100}) Transverse momenta of the leptons at \unit[100]{TeV} for a \unit[5]{TeV} heavy Higgs.
}
\label{fig:lepton pt}
\end{figure}

Given that gluon fusion and bottom quark associated production modes are unpromising, we propose to cover the low $\tan \beta$ region through top associated production modes, namely the channels
\begin{itemize}
\item $pp\to t\bar tH\to t\bar tt\bar t$ ,
\item $pp\to \bar tHW^+ b  \to \bar tt\bar tW^+ b\ $ and $\ pp\to tHW^- \bar b \to tt\bar tW^- \bar b\ $.%
\footnote{Whenever we will mention in the following just one of these two conjugated processes, we implicitly mean both $pp\to \bar tHW^+ b\ $ and $\ pp\to tHW^- \bar b\ $, together.}
\end{itemize}
These channels are ideally suited for probing the low $\tan\beta$ region, since they do not suffer significant interference with Standard Model backgrounds, and the corresponding production cross sections are maximized for low $\tan\beta$ in type~II 2HDM.
Various aspects of these channels have already been discussed~\cite{Han:2004zh, Lillie:2007hd, Acharya:2009gb, Chen:2015fca}.
In this work we develop an optimized strategy for probing extended Higgs sectors in top associated production by focusing on the three main kinematic features of these channels:
First, large heavy Higgs masses lead to a sizable scalar sum of transverse momenta $H_T$.
Second, we use the forward-/backwardness of the accompanying quarks.
Although the heavy Higgs is produced in association with one or two top quarks, the bottom quarks resulting from heavy Higgs decays still tend to be forward/backward, as shown in Figure~\ref{fig:delta eta}.
This feature discriminates against backgrounds other than the irreducible $t\bar tt\bar t$ background.
Third, we make use of the same-sign di-lepton (SSDL) signature present in these signal channels.
Especially in the $pp\to \bar tHW^+ b$ channel, the transverse momentum of the lepton originating from the decay of the associated W boson is comparable to the one originating from the decay of the heavy Higgs, as shown in Figure~\ref{fig:lepton pt}.

The paper is organized as follows:
In Section~\ref{sec:Higgs Sector}, we study general features of heavy Higgs associated production at the Large Hadron Collider and a future $pp$-collider operating at $\sqrt{s} = \unit[100]{TeV}$.
We then consider constraints from existing SSDL searches at the \unit[8]{TeV} LHC in Section~\ref{sec:constraints} in a simplified model framework with a single scalar or pseudoscalar heavy Higgs.
We then turn to prospects for probing both scalar and pseudoscalar heavy Higgses in type~II 2HDM at present and future colliders. We discuss the relevant backgrounds and introduce our analysis strategies in Section~\ref{sec:analysis}.
We present the results of these analyses in Section~\ref{sec:prospects} and we summarize our work in Section~\ref{sec:summary}. We reserve details of the BDT used in our \unit[14]{TeV} and \unit[100]{TeV} analyses for Appendix \ref{sec:boca tagger}, and a discussion of the related $bb$ associated production channel for Appendix \ref{sec:systematic error}.

\section{The Higgs Sector}\label{sec:Higgs Sector}

For the sake of concreteness, in this work we will focus on extended Higgs sectors whose low-energy physics can be characterized by a two Higgs doublet model of type~II, such as the Higgs sector of the MSSM.
In addition to the SM-like Higgs boson, these sectors contain two heavy neutral Higgs bosons --- one CP-even ($H$) and one CP-odd ($A$) --- and a pair of charged Higgs bosons ($H^\pm$).
The physics of these bosons is governed at tree level by a neutral mixing angle and the ratio of the Higgs vacuum expectation values $\tan\beta$.
In the particular case of the MSSM, the neutral mixing angle is further fixed by the mass scale of the heavy Higgs bosons. Current Higgs coupling measurements require such extended Higgs sectors to be near an \textit{alignment limit} of the parameter space~\cite{Gunion:2002zf, Craig:2012vn, Craig:2013hca, Carena:2013ooa, Haber:2013mia}, in which the couplings of the light CP-even Higgs scalar are Standard Model-like.

\subsection{Production of heavy scalars in association with top quark(s)}

\begin{figure}
\centering
\includegraphics[width=.75\textwidth]{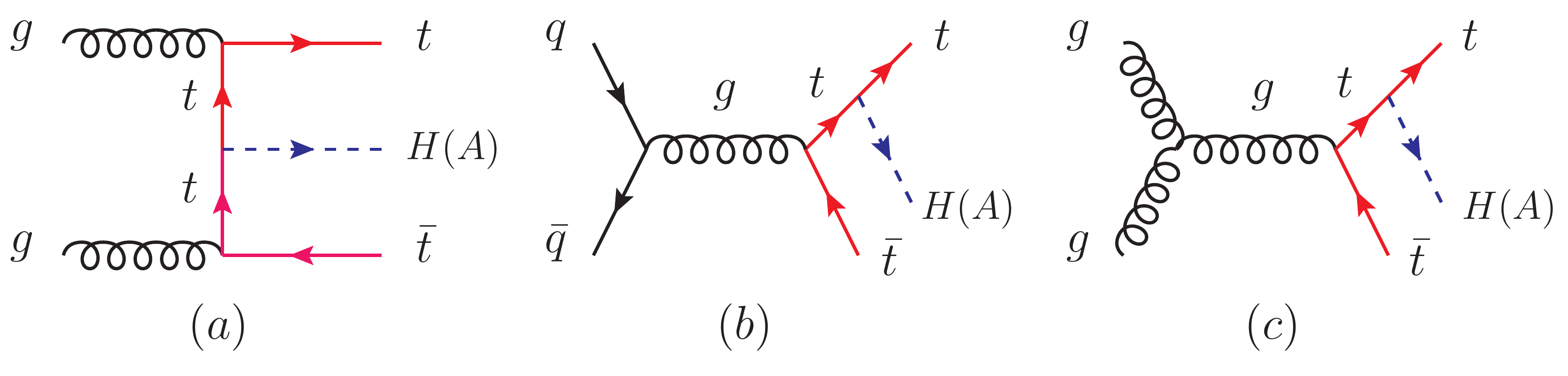}%
\sublabel{fig:4top_int}
\sublabel{fig:4top_ext_1}
\sublabel{fig:4top_ext_2}
\caption{Typical Feynman diagrams for the production of a heavy scalar in association with a top pair at proton-proton colliders.}
\label{fig:4top_feyn}
\end{figure}

The physics of the alignment limit provides a natural organizing principle for associated production modes of the heavy Higgs bosons.
In the alignment limit with small $\tan\beta$, the $HW^+W^-$ and $b\bar bH(A)$ couplings are suppressed, so that the dominant contributions to $H(A)$ production arise from the $t\bar tH(A)$ vertex.
This leads to a variety of production processes in association with $t \bar t$ pairs that can be generated from the standard model (SM) top production processes with an additional heavy scalar radiated from the internal top quark (Figure~\ref{fig:4top_int}) or an external top quark leg (Figure~\ref{fig:4top_ext_1} and~\ref{fig:4top_ext_2}).

\begin{figure}
\includegraphics[width=\textwidth]{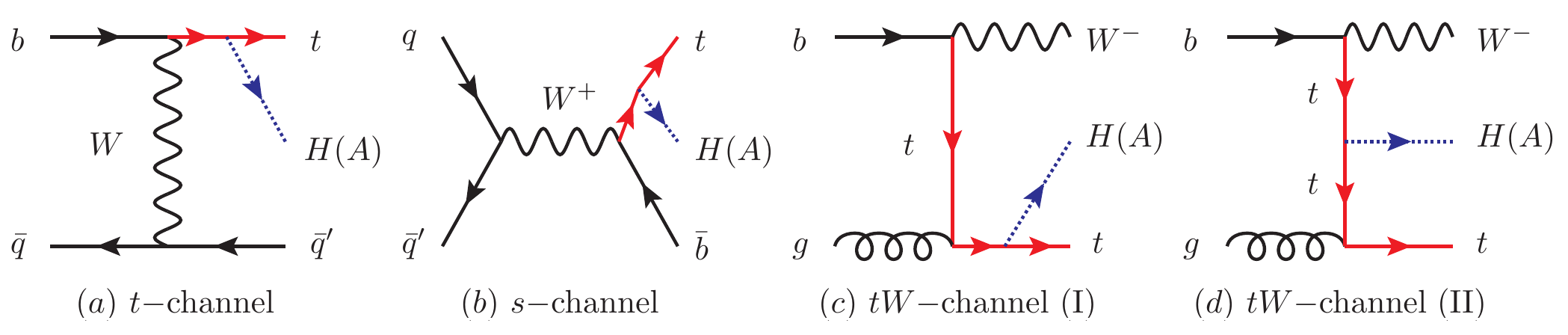}%
\sublabel{fig:1top_t}%
\sublabel{fig:1top_s}%
\sublabel{fig:1top_ext}%
\sublabel{fig:1top_int}%
\caption{Typical Feynman diagrams for the production of a heavy scalar in association with a single top quark at proton-proton colliders.}
\label{fig:1top_feyn}
\end{figure}

In addition to production of a heavy Higgs in association with top quark pairs, production in association with a single top quark may play a useful role. Production of Higgs bosons in association with single top quarks was studied extensively in~\cite{Maltoni:2001hu,Demartin:2015uha}, although the details differ somewhat near the alignment limit where radiation of heavy Higgses from vector bosons is suppressed.
The production of a heavy scalar in association with a single top quark in the alignment limit contains three main channels: $t$-channel (Figure~\ref{fig:1top_t}), $s$-channel (Figure~\ref{fig:1top_s}) and $tW$-associated production channel (Figure~\ref{fig:1top_ext} and~\ref{fig:1top_int}).
The $s$-channel process is highly suppressed by the center of mass energy $1/s^2$ and is much smaller than the other two.
Although the $t$-channel process is suppressed by a factor of $\alpha / \left( \alpha_s \sin^2 \theta_W \right)$, its cross-section is larger than that of the $tW$-associated channel on account of the larger phase space and the parton distribution function (PDF) of the valence quark.
However, as shown in Figure~\ref{fig:1top_feyn}, the cross-section of the $tW$-associated channel for the heavy scalar with single top production is increased by the additional possibility of internal radiation.
Furthermore, the suppression from the phase space volume is no longer significant when the scalar is heavy, because the volume of phase space is determined by the mass of the heavy scalar. Thus both the $t$-channel and $tW$-associated channels contribute significantly to the total cross section for production of a heavy Higgs boson in association with a single top quark.

Although a variety of search strategies are sensitive to this final state, in this work we focus on final states involving same-sign dileptons.
If we require the signal events to contain SSDL, the contribution from the $tWH(A)$ channel will be enhanced by the possibility of the charged lepton from the $W^\pm$ decay.
Hence we expect the dominant contributions from new physics in SSDL final states to come from the $tWH(A)$ channel, with a sub-dominant contribution coming from the $tqH(A)$ channel.
The $s$-channel contribution should be negligibly small as discussed above.

In comparison, the rate for $t\bar tH(A)$ production is only slightly suppressed by the phase space, and is enhanced relative to single-top processes by both the coupling constant of the strong interaction $\order{\alpha_s \sin^2 \theta_W / \alpha}$ and the gluon PDF.
Hence the contributions from $t \bar t$ associated production are expected to be significant, especially when searching for SSDL signals. In this work we consider both single-top and $t \bar t$ associated production processes.

\begin{figure}
\begin{subfigure}{.5\textwidth}
\includegraphics[width=\textwidth]{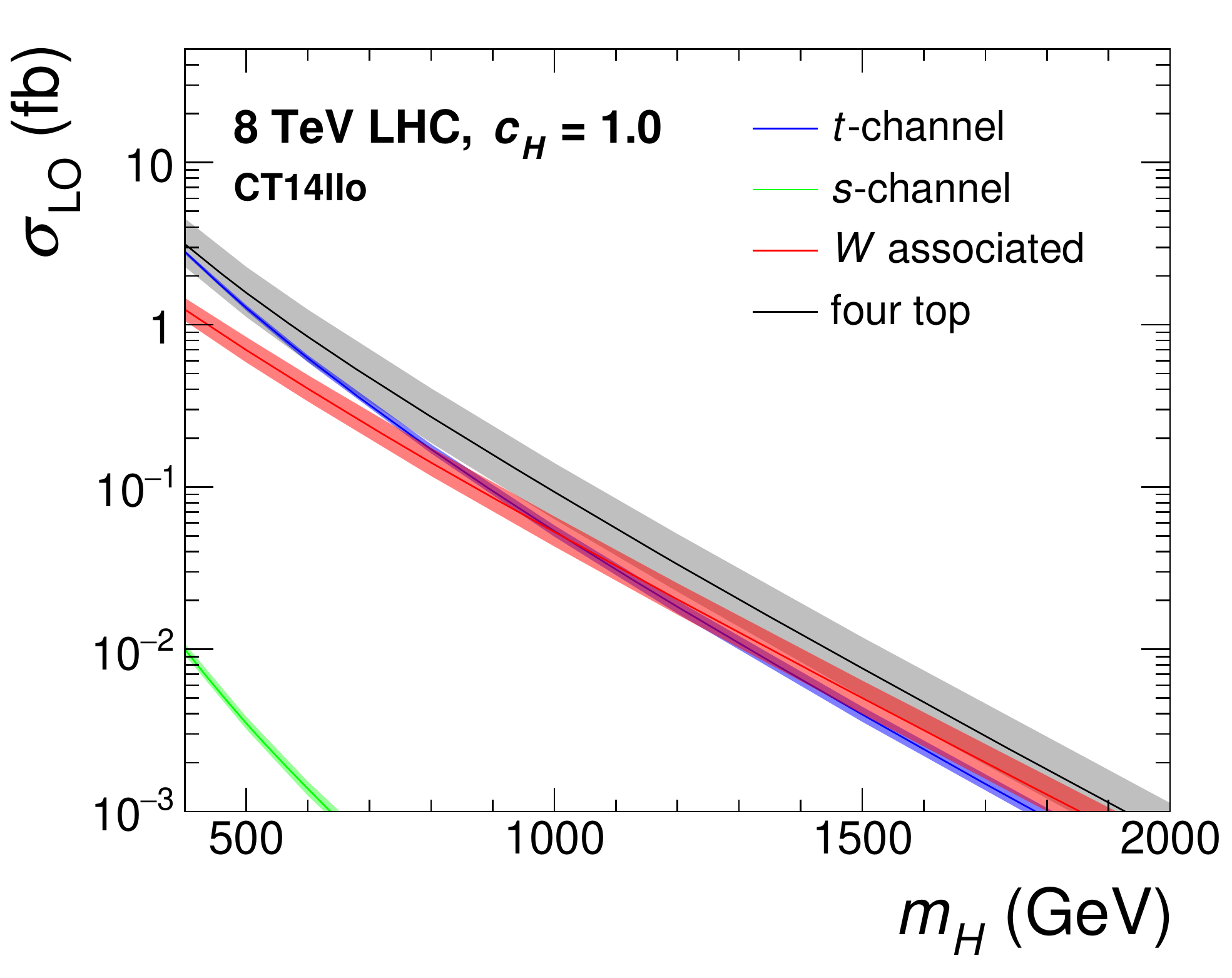}
\caption{CP-even heavy Higgs production at \unit[8]{TeV}}
\label{fig:even at 8 TeV}
\end{subfigure}
\begin{subfigure}{.5\textwidth}
\includegraphics[width=\textwidth]{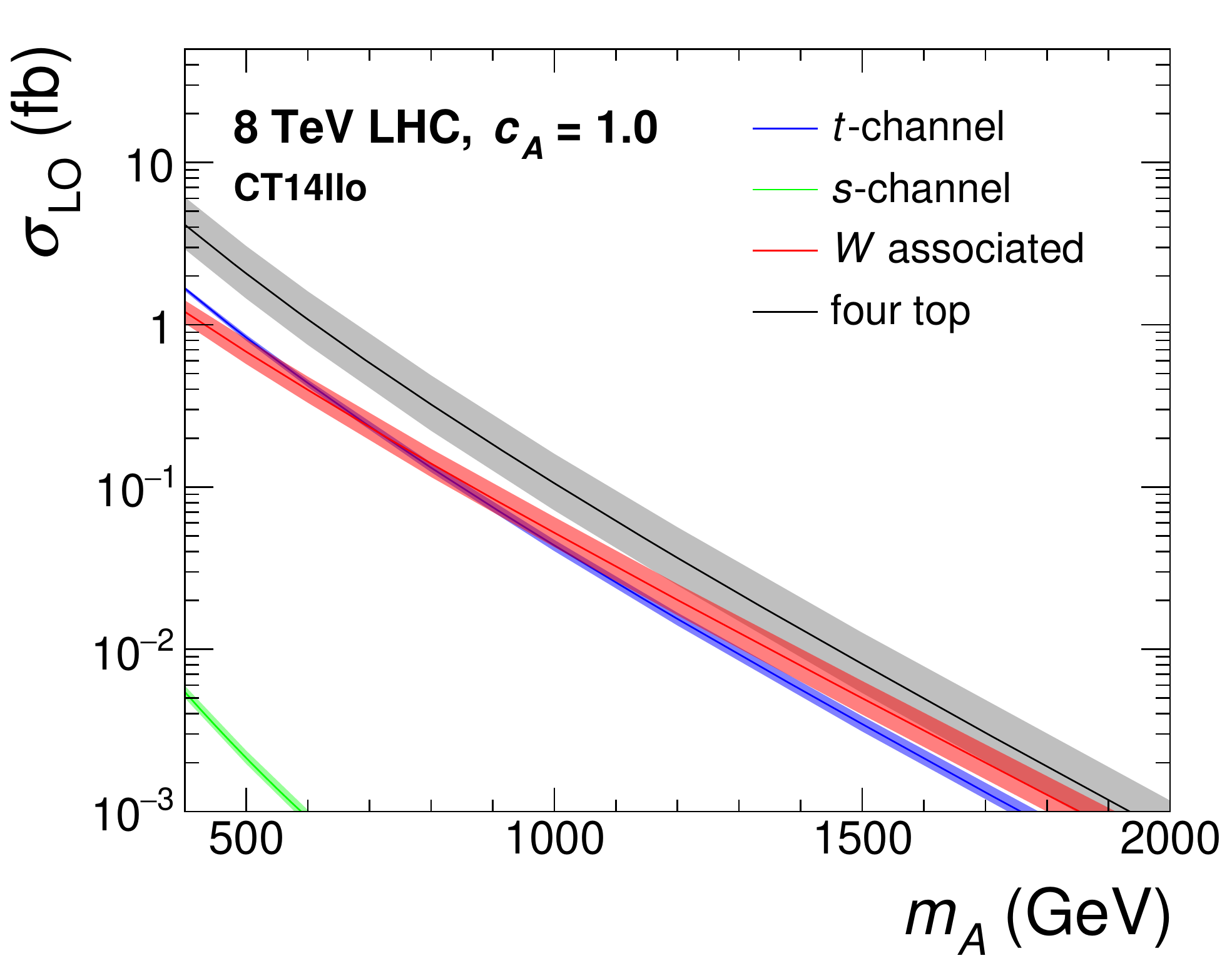}
\caption{CP-odd heavy Higgs production at \unit[8]{TeV}}
\label{fig:odd at 8 TeV}
\end{subfigure}
\begin{subfigure}{.5\textwidth}
\includegraphics[width=\textwidth]{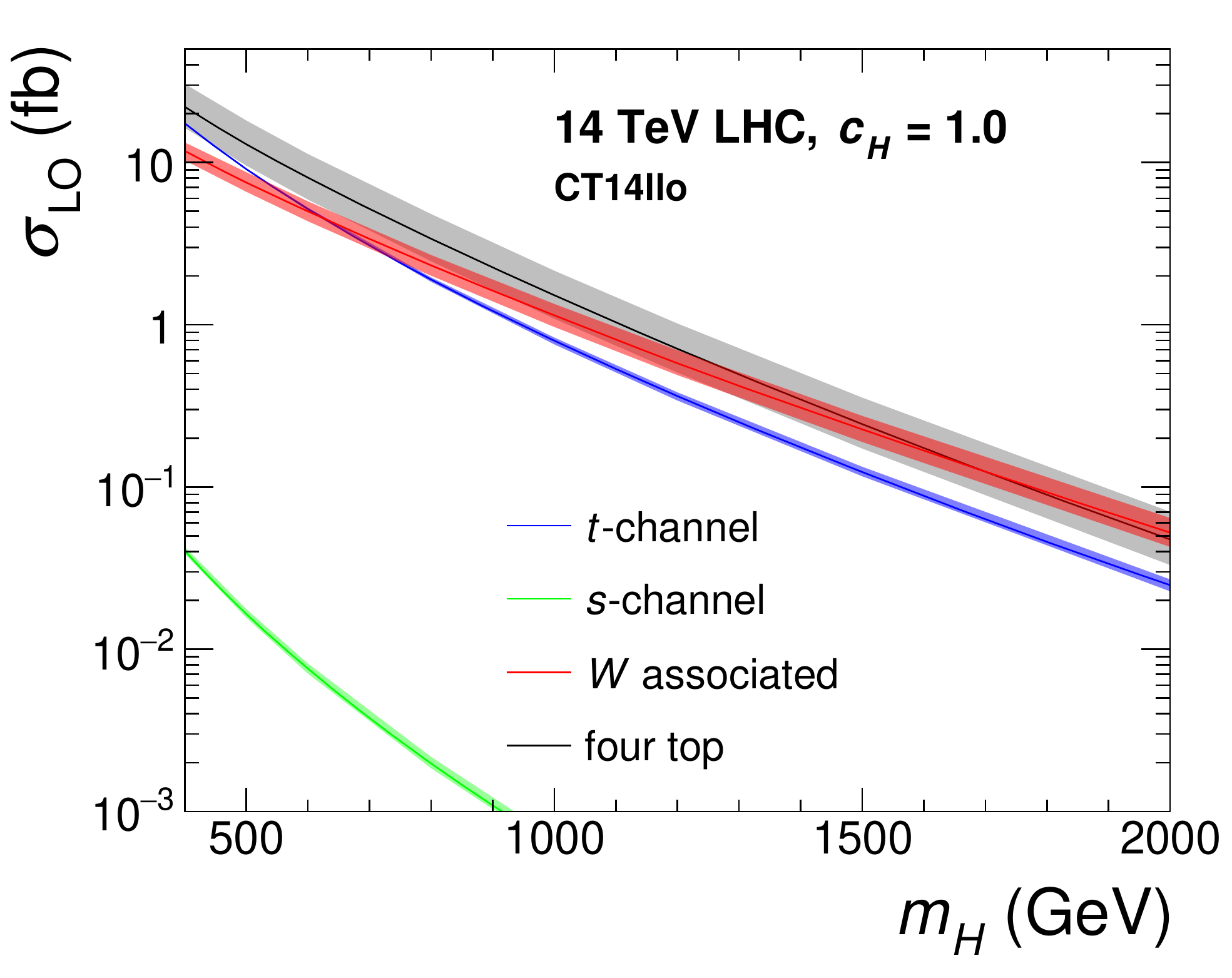}
\caption{CP-even heavy Higgs production at \unit[14]{TeV}}
\label{fig:even at 14 TeV}
\end{subfigure}
\begin{subfigure}{.5\textwidth}
\includegraphics[width=\textwidth]{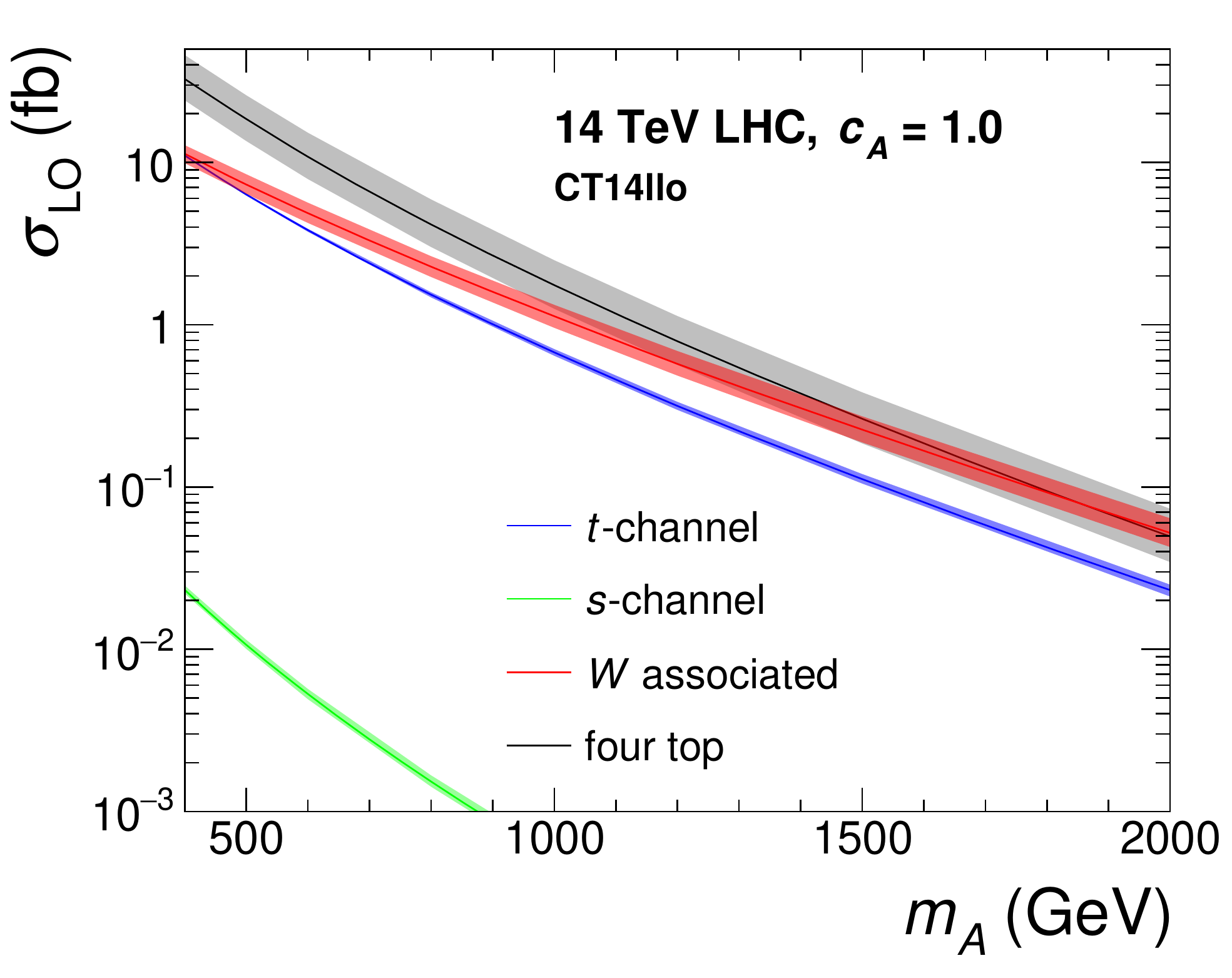}
\caption{CP-odd heavy Higgs production at \unit[14]{TeV}}
\label{fig:odd at 14 TeV}
\end{subfigure}
\caption{
(\subref{fig:even at 8 TeV}) Cross-sections of the 3-top (colored) and 4-top (black) processes for a CP-even heavy Higgs at the \unit[8]{TeV} LHC, where the number of top quarks includes both the associated states and the $t \bar t$ decay of the heavy Higgs.
(\subref{fig:odd at 8 TeV}) The corresponding cross sections for a CP-odd heavy Higgs.
(\subref{fig:even at 14 TeV}) Cross-sections of the 3-top (colored) and 4-top (black) processes for a CP-even heavy Higgs at the \unit[14]{TeV}.
(\subref{fig:odd at 14 TeV}) The corresponding cross sections for a CP-odd heavy Higgs.
In each case the shaded regions represent the scale uncertainties.
Note the difference of uncertainties for processes proportional to $\alpha_s$ and $\alpha_s^2$.
In simulating these cross-sections we have assumed that the heavy Higgs coupling to bottom quarks and $\tau$-lepton are negligible, which corresponds to small values of $\tan\beta$.
}
\label{fig:xsec_hao}
\end{figure}

\begin{figure}
\begin{subfigure}{.5\textwidth}
\includegraphics[width=\textwidth]{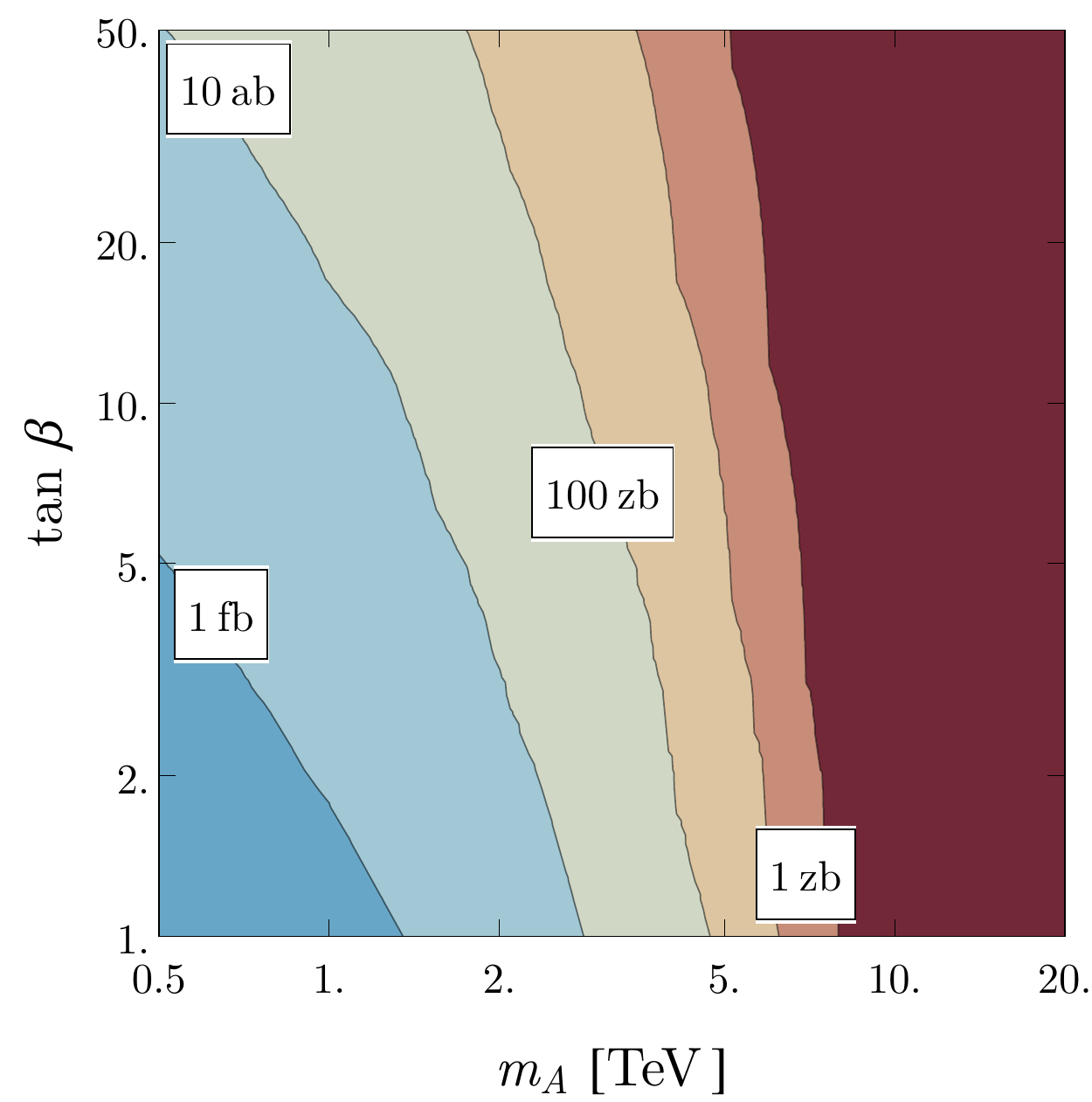}
\caption{$\sigma\left(pp\to (H+A)t\bar t\right)$ at \unit[14]{TeV}}
\label{fig:ProXLHCttHA}
\end{subfigure}
\hfill
\begin{subfigure}{.5\textwidth}
\includegraphics[width=\textwidth]{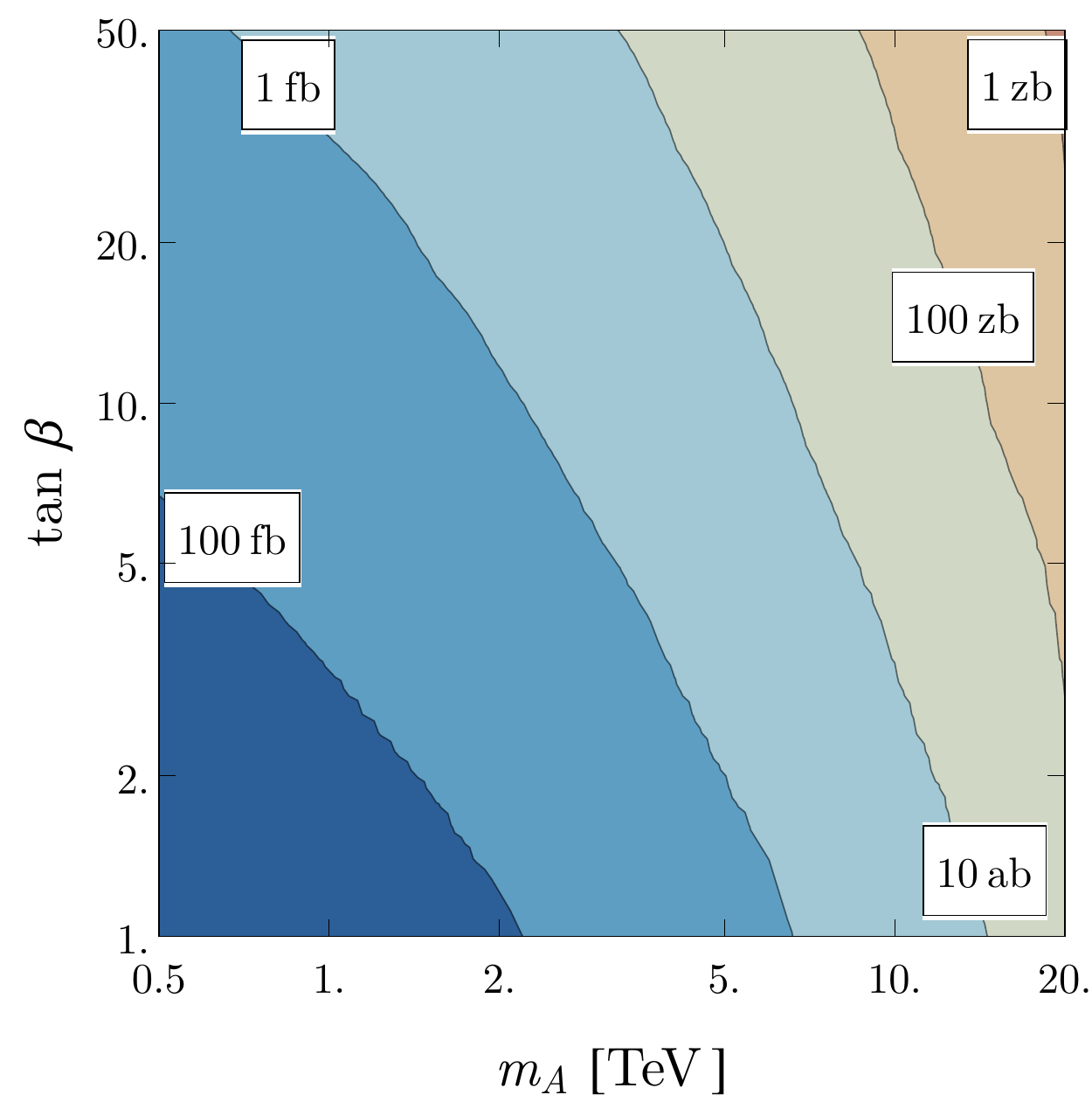}
\caption{$\sigma\left(pp\to (H+A)t\bar t\right)$ at \unit[100]{TeV}}
\label{fig:ProX100TeVttHA}
\end{subfigure}
\begin{subfigure}{.5\textwidth}
\includegraphics[width=\textwidth]{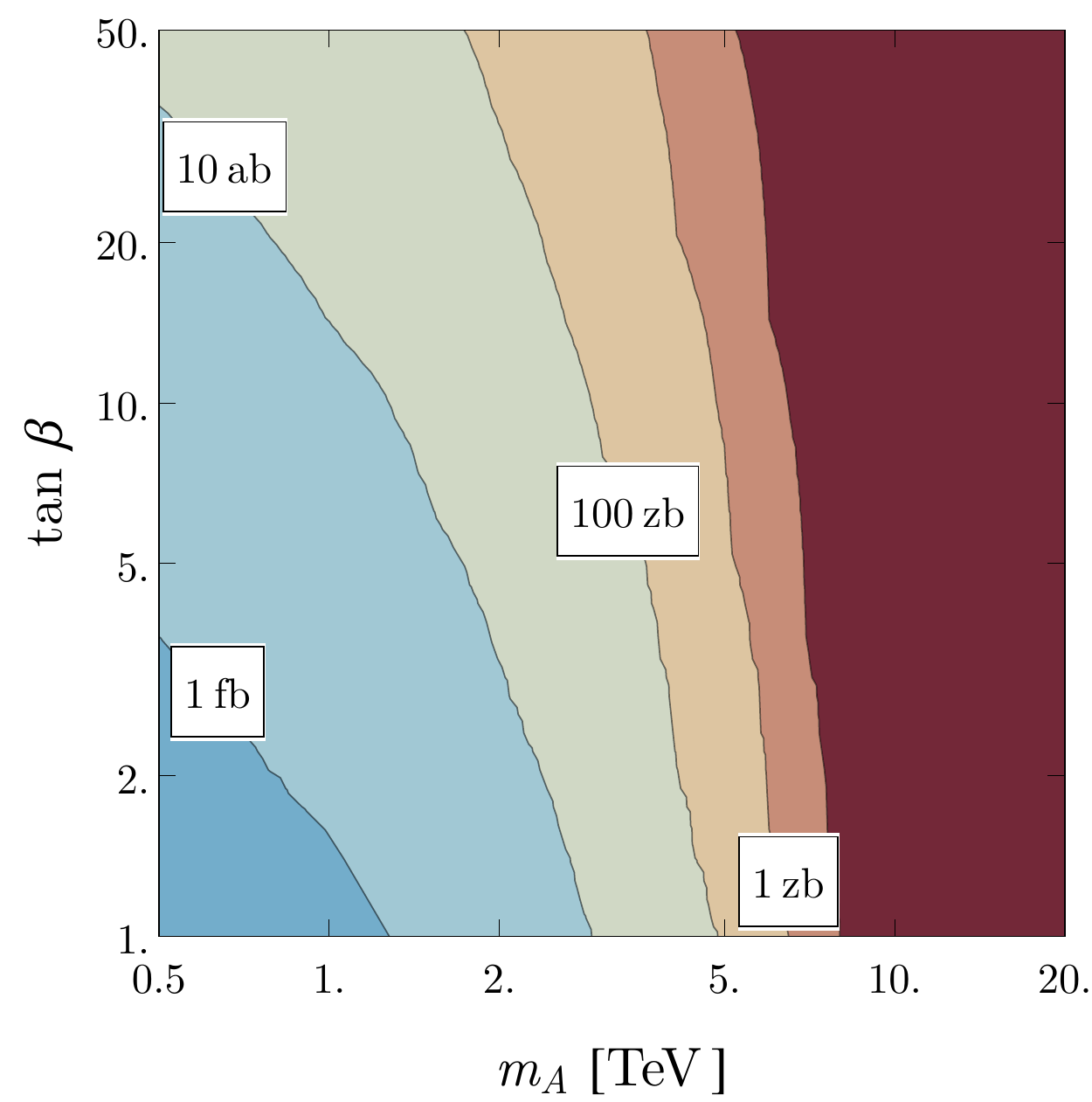}
\caption{$\sigma\left(pp\to (H+A)tW^\pm\right)$ at \unit[14]{TeV}}
\label{fig:ProXLHCtHA}
\end{subfigure}
\hfill
\begin{subfigure}{.5\textwidth}
\includegraphics[width=\textwidth]{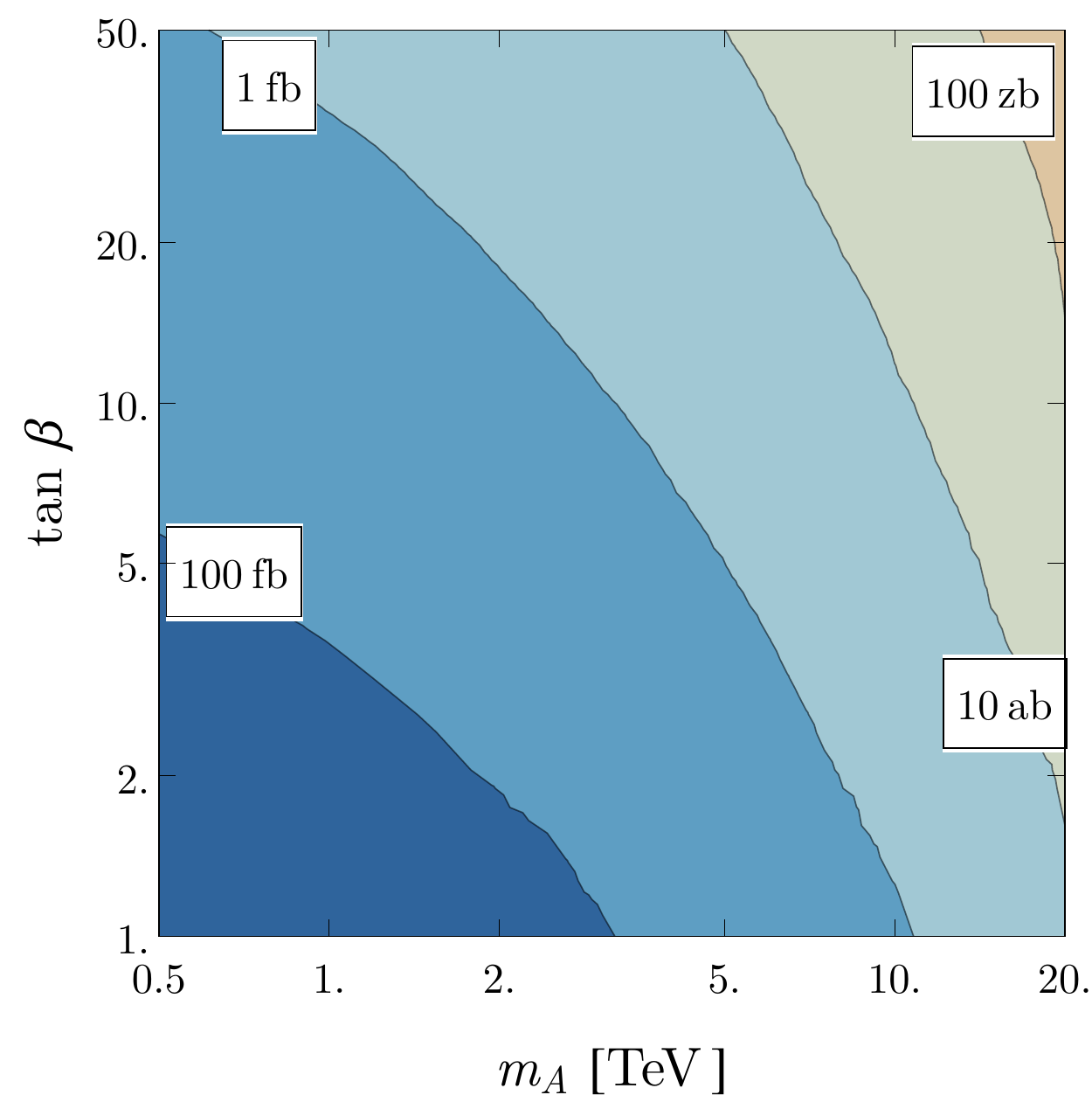}
\caption{$\sigma\left(pp\to (H+A)tW^\pm\right)$ at \unit[100]{TeV}}
\label{fig:ProX100TeVtHA}
\end{subfigure}
\caption{(\subref{fig:ProXLHCttHA})  Contours of the $t \bar t (H + A)$ associated production cross-section of heavy neutral Higgs bosons at the \unit[14]{TeV} LHC.
(\subref{fig:ProX100TeVttHA}) Contours of the $t \bar t (H + A)$ cross-sections at a \unit[100]{TeV} $pp$-collider.
(\subref{fig:ProXLHCtHA}) Contours of the $tW (H + A)$ associated production cross-section at the \unit[14]{TeV} LHC.
(\subref{fig:ProX100TeVtHA}) Contours of the $tW (H + A)$ cross-sections at a \unit[100]{TeV} $pp$-collider.
The cross-section are calculated to leading order with \software{MadGraph} using its variable factorization and renormalization scale.
}
\label{fig:ProductionXsectiont}
\end{figure}

In what follows, we will both obtain existing limits on these processes by reinterpreting SSDL searches at $\sqrt{s} = \unit[8]{TeV}$ and forecast the reach of the $\sqrt{s} = \unit[14]{TeV}$ LHC and future $pp$-collider in SSDL channels.
To do so, we work in terms of a simplified model in which $H(A)$ couples to the SM particles via
\begin{align}
    \mathcal L
  = - y_t ( c_H H \bar t t + i c_A A \bar t \gamma_5 t )
\ , \label{eq::CP-structure}
\end{align}
where $y_t$, $y_b$ and $y_\tau$ are the SM Yukawa coupling constant of the third generation leptons.
As we are focusing on the case with small $\tan\beta$ we will neglect the sub-dominant coupling to $b$ and $\tau$ when we derive limits on the coefficients $c_H, c_A$.

We calculate the leading order (LO) cross-sections using \software[5]{MadGraph}~\cite{Alwall:2014hca} with CT14llo PDF in the 5-flavor scheme (FS)~\cite{Dulat:2015mca}.
For the $tH(A)+X$ processes, we choose the factorization and renormalization scales to be $\mu_F = \mu_R = m_t^{\overline{\text{MS}}} + m_{H(A)}$
, where
$m_t^{\overline{\text{MS}}} = \unit[163]{GeV}$
is the $\overline{{\text{MS}}}$ mass of the SM top quark.
For $t\bar tH(A)$ process, we instead choose
$\mu_F = \mu_R = m_t^{\overline{\text{MS}}} + m_{H(A)} / 2$.
The resulting cross sections for the \unit[8]{TeV} and \unit[14]{TeV} LHC are shown in Figure~\ref{fig:xsec_hao}.
Notice that the 4-top process has larger scale uncertainty since it is $\order{\alpha_s^2}$, while the 3-top processes are $\order{\alpha_s}$.

For the \unit[14]{TeV} LHC and a future \unit[100]{TeV} $pp$-collider we present additionally the contours in the $m_A$-$\tan\beta$ plane for the production cross-sections of the $pp\to t\bar tH(A)$ and the $pp\to tWH(A)$ processes within the framework of the MSSM in Figure~\ref{fig:ProductionXsectiont}.
We point out that for low masses the cross-section of the $tWH(A)$ processes is smaller than that of $t\bar tH(A)$ processes, but that the situation reverses at large heavy Higgs masses due to the asymptotic freedom of $\alpha_s$ together with the faster falloff in $x$ of the gluon PDF relative to the bottom-quark PDF.

\section{Constraint from LHC Run I} \label{sec:constraints}

\begin{figure}
\centering
\includegraphics[width=.6\textwidth]{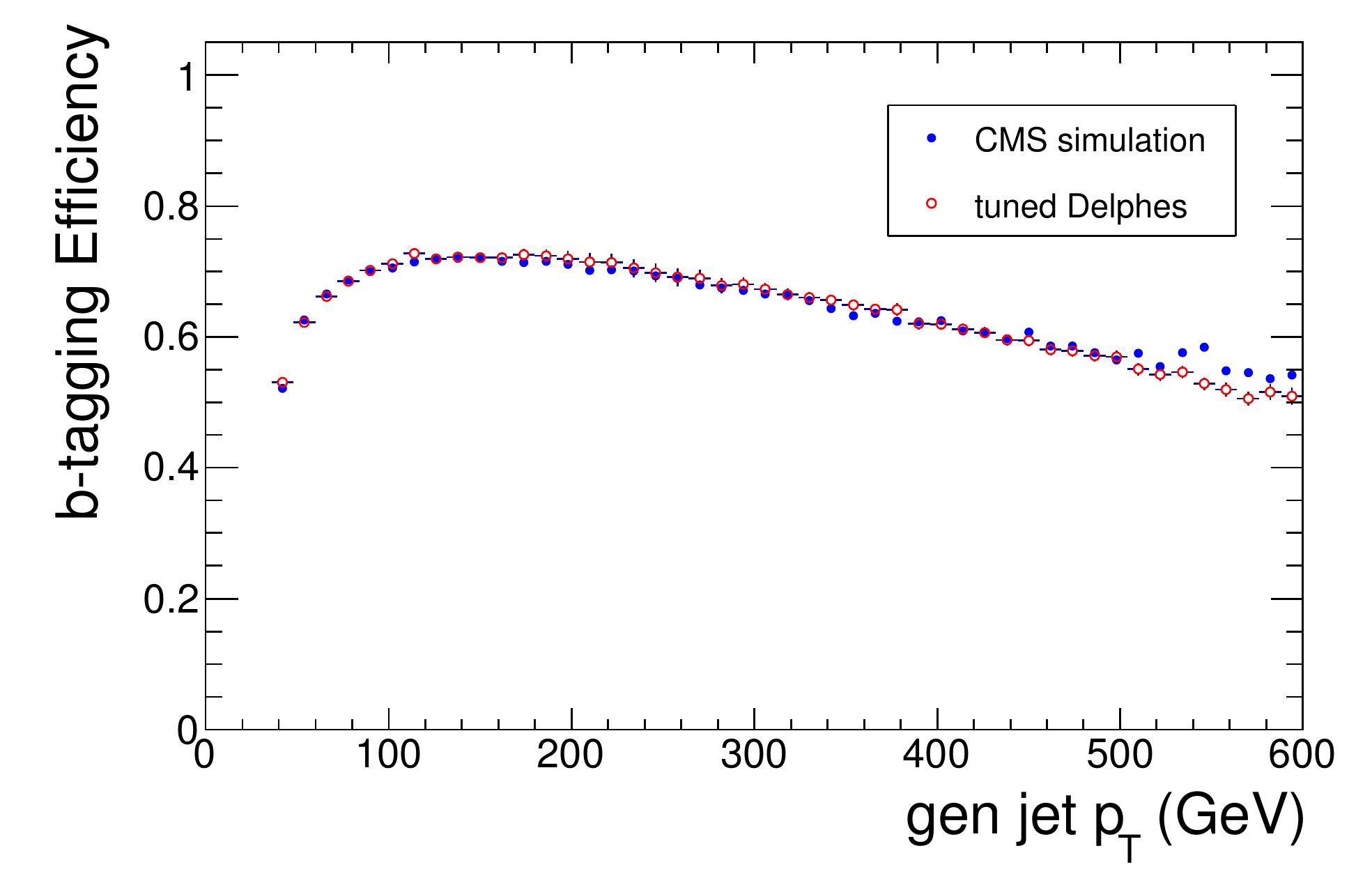}
\caption{
$b$-tagging efficiency as a function of  jet $p_T$ at the \unit[8]{TeV} LHC.
The red hollow points are the result from our simulation using $pp\to Zb(\bar b)$ events.
The solid blue points are the result from the CMS simulation presented in~\cite{Chatrchyan:2013fea}. Although~\cite{Chatrchyan:2013fea} uses simulated $t \bar t$ events, the $b$-tagging efficiency is not expected to differ substantially between different processes.}
\label{fig:btag}
\end{figure}

Before developing a strategy for future searches, it is useful to determine the state of existing limits from searches at the \unit[8]{TeV} LHC.
In this section, we find the constraint on heavy Higgs bosons coming from an SSDL search performed by the CMS collaboration~\cite{Chatrchyan:2013fea}.
We consider a simplified model containing a heavy scalar or pseudoscalar Higgs boson with coupling to the top quark as in Equation \eqref{eq::CP-structure}.
The three-top signal events are generated to leading order (LO) at parton level with CTEQ6L1 PDFs~\cite{Pumplin:2002vw} within the 5-flavor scheme using \software{MadGraph}~\cite{Alwall:2014hca}.
The renormalization and factorization scales are set to be $\mu_R = \mu_F = m_t + m_{H(A)}$.
We also generate four-top signal events with the same method as the three-top signal events.
The renormalization and factorization scales for four-top processes are set to be $\mu_R = \mu_F = m_t + m_{H(A)} / 2$.

In both cases, the top quarks in the final state are decayed using \software{MadGraph} in order to preserve the effects of spin correlations. Events are showered using \software[6.4]{PYTHIA}~\cite{Sjostrand:2006za} with the Z2 Tune~\cite{Field:2011iq}.
\software[3]{Delphes}~\cite{deFavereau:2013fsa, Cacciari:2011ma} is used to simulate the detector effects.
The $b$-tagging efficiency from \software{Delphes} is tuned to mimic the results of the simulation by the CMS collaboration (cf. Figure~\ref{fig:btag}).
The mis-tagging rate of charm jets (light jets) is \unit[20]{\%} (\unit[1]{\%})~\cite{Chatrchyan:2013fea}.
We check the cut acceptances of events with same-sign top-quark pair production.
The same-sign top search regions are defined as having $\geq 2$ jets, $E_T^\text{miss} > \unit[30]{GeV}$ and $H_T > \unit[80]{GeV}$.
Additionally, the number of $b$-jets has to be either equal to one (SStop1) or greater than one (SStop2). The SStop1 (SStop2) search region acceptance (including branching fractions) is \unit[0.43]{\%} (\unit[0.26]{\%}) with relative uncertainty \unit[14]{\%} from simulation by the CMS collaboration~\cite{Chatrchyan:2013fea}, while our result is \unit[0.39]{\%} (\unit[0.28]{\%}). The results of our simulation are consistent with the results given by the CMS collaboration to within the Monte Carlo uncertainty.

\begin{table}
\centering
\begin{tabular}{cccccccccc}
    \toprule
    $N_{b\text{-jets}}$
  & $p_T(l)$ [GeV]
  & \multicolumn{8}{c}{$E_T^\text{miss}$ [GeV]}
 \\ \cmidrule{3-10}
  &
  & \multicolumn{4}{c}{50--120}
  & \multicolumn{4}{c}{$> 120$}
 \\
  &
  & \multicolumn{4}{c}{$H_T$ [GeV]}
  & \multicolumn{4}{c}{$H_T$ [GeV]}
 \\ \cmidrule(r){3-6} \cmidrule(l){7-10}
  &
  & \multicolumn{2}{c}{200--400}
  & \multicolumn{2}{c}{$>400$}
  & \multicolumn{2}{c}{200--400}
  & \multicolumn{2}{c}{$>400$}
 \\
  &
  & \multicolumn{2}{c}{$N_\text{jets}$}
  & \multicolumn{2}{c}{$N_\text{jets}$}
  & \multicolumn{2}{c}{$N_\text{jets}$}
  & \multicolumn{2}{c}{$N_\text{jets}$}
 \\ \cmidrule(r){3-4} \cmidrule(lr){5-6} \cmidrule(lr){7-8} \cmidrule(lr){9-10}
  &
  & 2--3
  & $\geq4$
  & 2--3
  & $\geq4$
  & 2--3
  & $\geq4$
  & 2--3
  & $\geq4$
 \\ \midrule \multirow{3}{*}[-1.1ex]{0}
  &
  & SR01
  & SR03
  & SR02
  & SR04
  & SR05
  & SR07
  & SR06
  & SR08
 \\ \cmidrule(r){3-4} \cmidrule(lr){5-6} \cmidrule(lr){7-8} \cmidrule(lr){9-10}
  & $>10$
  & 22
  & 9.6
  & 15
  & 3.2
  & 12
  & 3.3
  & 15
  & 4.2
 \\
  & $>20$
  & 13
  & 4.0
  & 10
  & 2.8
  & 4.4
  & 2.8
  & 10
  & 4.4
 \\ \midrule \multirow{3}{*}[-1.1ex]{1}
  &
  & SR11
  & SR13
  & SR12
  & SR14
  & SR15
  & SR17
  & SR16
  & SR18
 \\ \cmidrule(r){3-4} \cmidrule(lr){5-6} \cmidrule(lr){7-8} \cmidrule(lr){9-10}
  & $>10$
  & 22
  & 7.5
  & 5.1
  & 4.0
  & 4.9
  & 4.0
  & 5.4
  & 12
 \\
  & $>20$
  & 7.5
  & 4.0
  & 7.1
  & 2.8
  & 8.0
  & 5.1
  & 3.2
  & 9.7
 \\ \midrule \multirow{3}{*}[-1.1ex]{$\geq 2$}
  &
  & SR21
  & SR23
  & SR22
  & SR24
  & SR25
  & SR27
  & SR26
  & SR28
 \\ \cmidrule(r){3-4} \cmidrule(lr){5-6} \cmidrule(lr){7-8} \cmidrule(lr){9-10}
  & $>10$
  & 10
  & 5.4
  & 3.1
  & 15
  & 2.7
  & 2.0
  & 5.0
  & 4.5
 \\
  & $>20$
  & 13
  & 4.3
  & 3.5
  & 11
  & 6.5
  & 2.0
  & 3.7
  & 4.1
 \\ \bottomrule
\end{tabular}
\caption{The \unit[2]{$\sigma$} upper bounds of the number of signal events for each signal region (SR) defined in~\cite{Chatrchyan:2013fea}.}
\label{table:cms8tev_stat}
\end{table}

Given $n$ observed events the significance is~\cite{Cowan:2010js}
\begin{equation}
    Z(x | n)
  = \sqrt{- 2 \ln \frac{L (x | n )}{L (n | n )}}
\ , \label{eq:significance}
\end{equation}
where $x$ is either the number of events predicted by the background only hypothesis $b$ or by the background with signal hypothesis $b+s$ and the likelihood function is given by the Poisson probability
\begin{equation}
    L (x | n)
  = \frac{ x^n e^{-x}}{n!}
\ .
\end{equation}
For the exclusion of a model we require $Z(b + s | n) \geq 2$ and for discovery we require $Z(b | n) \geq 5$.%
\footnote{%
In the case that $n < b$ we calculate the significance with the more conservative statistics also used at the Tevatron, namely $Z = \sqrt{- 2 \ln \frac{L (b + s | n )}{L (b | n )}}$.
This split approach corresponds to the test statistic $\tilde q_\mu$ in~\cite{Cowan:2010js}.}
For the projection to future experiments we replace the event number $n$ with the prediction for the alternative hypothesis.
Hence, we are using $Z(b + s | b) \geq 2$ and $Z(b | b+s) \geq 5$, for exclusion and discovery, respectively.
Here we assume positive interference between signal and background and the systematic uncertainties of the signal and the background estimation are not included in this simplified statistical procedure.
Using the result in Table~7 of~\cite{Chatrchyan:2013fea}, we can give the \unit[2]{$\sigma$} upper bound of the event numbers of the new physics model for each signal region defined in~\cite{Chatrchyan:2013fea}.
The results are summarized in Table~\ref{table:cms8tev_stat}.

Given a number of signal events $s$ with $c_X = 1$ (where $X = H, A$) and an observed upper bound $n$ in the same signal region, the central value of the upper bound on $c_X$ is given by
\begin{equation}
   c_X
 = \sqrt\frac ns
\ .
\end{equation}
The corresponding error on the bound on $c_X$ is
\begin{equation}
    \delta c_\pm
  = \frac{\delta n}{2\sqrt{ns}}=\frac{c_X}{2\sqrt n}
\ .
\end{equation}

\begin{table}
\centering
\begin{tabular}{cccccc}
\toprule
    Mass [GeV]
  &
  & $t$-channel
  & $s$-channel
  & $W$ associated
  & 4-top
 \\ \midrule
    \multirow{2}{*}{400}
  & $c_H$
  & $10.3 \pm 2.43$
  & $91.7 \pm 21.6$
  & $7.98 \pm 1.87$
  & $4.47 \pm 1.06$
 \\
  & $c_A$
  & $9.63 \pm 2.27$
  & $123 \pm 29.0$
  & $8.40 \pm 1.98$
  & $4.28 \pm 1.01$
 \\ \midrule
    \multirow{2}{*}{500}
  & $c_H$
  & $9.26 \pm 2.19$
  & $130 \pm 30.7$
  & $9.55 \pm 2.24$
  & $5.81 \pm 1.37$
 \\
  & $c_A$
  & $11.4 \pm 2.69$
  & $163 \pm 38.4$
  & $9.89 \pm 2.33$
  & $5.37 \pm 1.27$
 \\ \midrule
    \multirow{2}{*}{600}
  & $c_H$
  & $12.0 \pm 2.83$
  & $186 \pm 43.9$
  & $12.0 \pm 2.82$
  & $7.55 \pm 1.78$
 \\
  & $c_A$
  & $19.9 \pm 4.69$
  & $228 \pm 53.5$
  & $12.3 \pm 2.90$
  & $6.99 \pm 1.65$
 \\ \midrule
    \multirow{2}{*}{800}
  & $c_H$
  & $20.5 \pm 4.83$
  & $531 \pm 125$
  & $19.5 \pm 4.56$
  & $12.7 \pm 3.00$
 \\
  & $c_A$
  & $23.6 \pm 5.55$
  & $441 \pm 104$
  & $19.6 \pm 4.62$
  & $11.9 \pm 2.81$
 \\ \midrule
    \multirow{2}{*}{1000}
  & $c_H$
  & $36.2 \pm 8.55$
  & $760 \pm 179$
  & $32.1 \pm 7.52$
  & $21.5 \pm 5.05$
 \\
  & $c_A$
  & $40.0 \pm 9.45$
  & $848 \pm 200$
  & $32.1 \pm 7.55$
  & $20.2 \pm 4.76$
 \\ \bottomrule
\end{tabular}
\caption{The \unit[2]{$\sigma$} upper bounds of the effective interaction strength $c_H$ and $c_A$ from \unit[8]{TeV} LHC.}
\label{table:cms8tev_result}
\end{table}

\begin{figure}
\begin{subfigure}{0.5\textwidth}
\includegraphics[width=\textwidth]{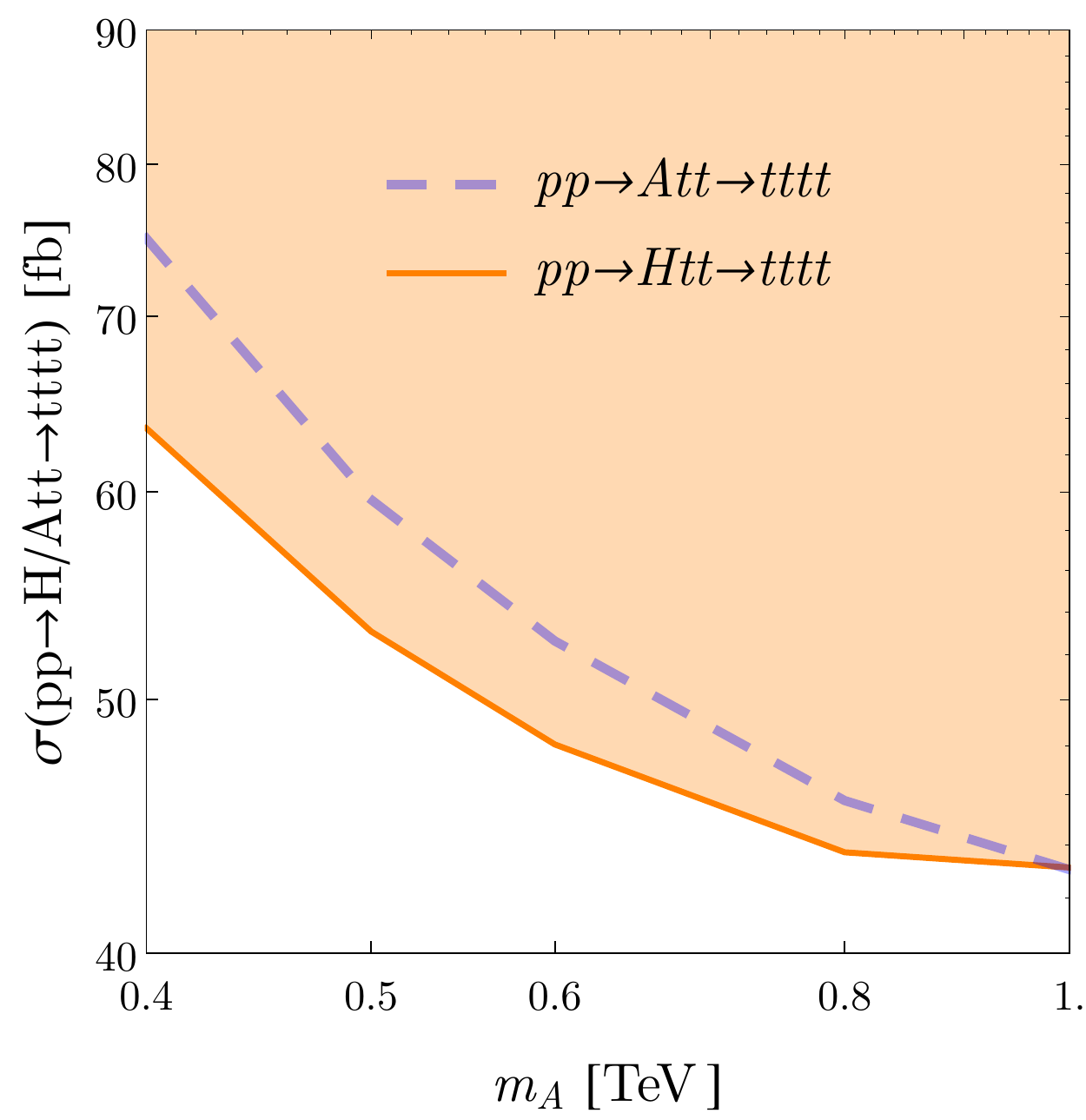}
\caption{4-top}
\label{fig:fourtop}
\end{subfigure}
\begin{subfigure}{0.5\textwidth}
\includegraphics[width=\textwidth]{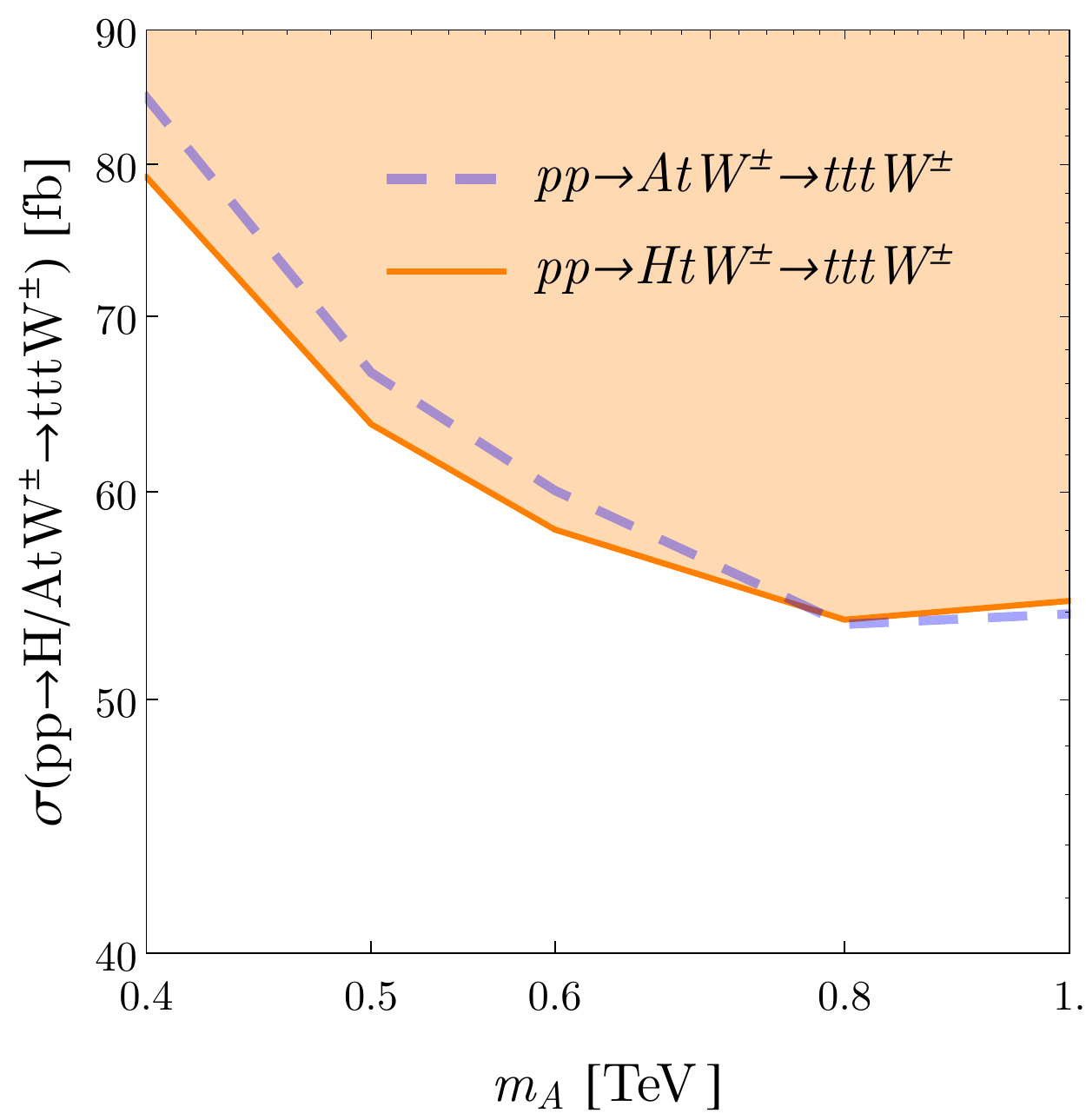}
\caption{3-top}
\label{fig:threetop}
\end{subfigure}
\caption{%
(\subref{fig:fourtop}) Exclusion cross section for  $pp\to H/Att \to tttt$  using \unit[8]{TeV} LHC data. (\subref{fig:threetop}) Exclusion cross section for  $pp\to A/H tW^\pm\to tttW^\pm$  using \unit[8]{TeV} LHC data.}
\label{fig:cms8tev_result}
\end{figure}

We check all of the 24 signal regions (SR01--SR08, SR11--SR18, SR21--SR28) for $H$ and $A$ with $m_{H(A)} = \unit[400, 500, 600, 800, 1000]{GeV}$.
The numerical simulation indicates that the strongest constraint is from the ``Low-$p_T$ SR28'' signal region for all of the mass points.
For each mass point we consider the contributions from the $t$-channel, $s$-channel, $W$ associated production channel and 4-top channel separately and in combination.
As shown in Table~\ref{table:cms8tev_result},  the strongest constraint is from the 4-top channel as expected. The combined results are shown in Figure~\ref{fig:cms8tev_result}.
From this it is apparent that \unit[8]{TeV} results place no meaningful limit on heavy scalar or pseudoscalar Higgses decaying to $t \bar t$ with $c_H, c_A \lesssim 4$.

\section{Analysis strategies for future searches} \label{sec:analysis}

\begin{table}
\centering
\begin{tabular}{rcccc}
    \toprule
    Background
  & \multicolumn{2}{c}{\unit[14]{TeV}}
  & \multicolumn{2}{c}{\unit[100]{TeV}}
 \\ \cmidrule(r){2-3} \cmidrule(l){4-5}
  & $\sigma$ [fb]
  & $\mathcal L_\text{gen}~[\unit{ab^{-1}}]$
  & $\sigma$ [fb]
  & $\mathcal L_\text{gen}~[\unit{ab^{-1}}]$
 \\ \midrule
    $t\bar t t\bar t$
  & 0.4851
  & 103
  & 122.7
  & 1.63
 \\ $tttW^\pm b$
  & 0.06016
  & 831
  & 14.86
  & 6.73
 \\ $t\bar tW^{+} b\bar b$
  & 0.03284
  & 1520
  & 0.3822
  & 262
 \\ \bottomrule
\end{tabular}
\caption{%
Leading order cross-section times branching ratio to same-sign dileptons for the dominant backgrounds at \unit[14]{TeV} and \unit[100]{TeV}, defined such that $tttW^\pm b$ does not have an overlapping contribution with $t\bar t t\bar t$.
We have checked that $t\bar tW^- b\bar b$ is sub-dominant to $t\bar tW^+ b\bar b$ at $pp$ colliders and will neglect it in the further discussion.
}
\label{tab:XSectionBG}
\end{table}

We next consider the prospects for optimized searches at the \unit[14]{TeV} LHC and a future \unit[100]{TeV} $pp$ collider. We consider both a cut-based analysis and a BDT analysis at \unit[14]{TeV}, the former providing a validation of the latter. For our \unit[100]{TeV} projections we consider only a BDT analysis for simplicity, given the considerable uncertainty regarding the parameters of a \unit[100]{TeV} collider. For all of these analyses we generate events in the four FS using \software[2.3.3]{MadGraph}, such that the $tW^\pm H(A)$ signal is generated with an additional soft and forward/backward $b$-quark.
We consider only events with jets harder than \unit[20]{GeV} for the LHC and \unit[40]{GeV} a future collider, respectively. We have kept the spin correlation using \software{MadSpin}.

We generate the dominant irreducible backgrounds $t\bar t t\bar t$ and  $tttW^\pm b$, as well as the main reducible background $t\bar tW^{+}bb$.
Given that our main analysis strategy involves SSDL, a mass-window cut around the $Z$ peak, at least four b-jets and a veto on the third lepton,
we have verified that the background contributions from $t\bar t W^\pm Z$, $t\bar tW^\pm cc$, $t\bar tZ$, $t\bar th$ and $ttW^\pm$ are sub-dominant to the main backgrounds.
The generated processes, their cross-sections, and the generated luminosity are summarized in Table~\ref{tab:XSectionBG}.

We use \software[3]{Delphes} to simulate a detector with CMS geometry.
For the \unit[14]{TeV} LHC, we use a tracker coverage of $|\eta| < 2.5$.
We assume a tracker coverage of $|\eta| < 3.5$ for detectors at future \unit[100]{TeV} colliders.
For all analyses we use anti-$k_T$ jet clustering with a jet cone size of 0.5.
However, in the BDT-based analysis this parameter plays a tangential role, as we additionally consider exclusively defined sub-jets as well as objects built of multiple jets.

\subsection{Pre-selection cuts}

For the analyses covering the \unit[14]{TeV} LHC we apply the following pre-selection cuts:
We consider only events with exactly two leptons with identical charge and transverse momenta $p_T > \unit[15]{GeV}$.
The leptons must be isolated with an isolation radius of $\Delta R > 0.3$ and a maximal transverse momentum ratio of 0.2.
Where the transverse momentum ratio is defined between the lepton and other cell activity within the isolation cone.
If the lepton transverse momentum is larger than \unit[50]{GeV}, we do not require them to be isolated~\cite{Brust:2014gia}.
We veto on a third lepton with $p_T > \unit[10]{GeV}$.
Additionally we require a minimal missing transverse energy $E_T^\text{miss} > \unit[30]{GeV}$ and reject events with less than four jets.

For the \unit[100]{TeV} analysis we demand for the leading lepton a transverse momentum of \unit[100]{GeV}.
The second leading lepton must have a $p_T>\unit[50]{GeV}$ and we veto on a third lepton with $p_T>\unit[50]{GeV}$.
We do not require leptons to be isolated as long as they are harder than \unit[100]{GeV}.
Otherwise, we apply the same isolation cone parameter as in the case of the LHC.
The missing transverse energy must be $E_T^\text{miss} > \unit[60]{GeV}$.

\subsection{Cut-based analysis for \unit[14]{TeV}} \label{sec:cut-based}

For the cut based analysis we use the \software{BoCA} $b$-tag presented in Appendix~\ref{sec:boca tagger} and require a minimum $b$-jet $p_T$ of \unit[40]{GeV}.
For this analysis we follow the ideas in~\cite{Aad:2015gdg}.
We require at least four bottom jets and veto events if the invariant mass of an electron pair falls within a \unit[20]{GeV} window around the $Z$-peak.
We develop a cut strategy by using the rectangular cut optimization of the \software{TMVA} library~\cite{Hocker:2007ht}.
The dominant cut is on the scalar sum of transverse momenta $H_T$.
In order to probe a given Higgs mass $m_{H/A}$ we require $H_T > m_{H/A}$.
These cuts also ensure the suppression of instrumental backgrounds, especially the misidentification of jets as well as the misidentification of lepton charges~\cite{ATLAS:2016sno}.

\subsection{BDT analysis for \unit[14 and 100]{TeV}} \label{sec:BDT}

\begin{figure}
\centering
\includegraphics[width=.7\textwidth]{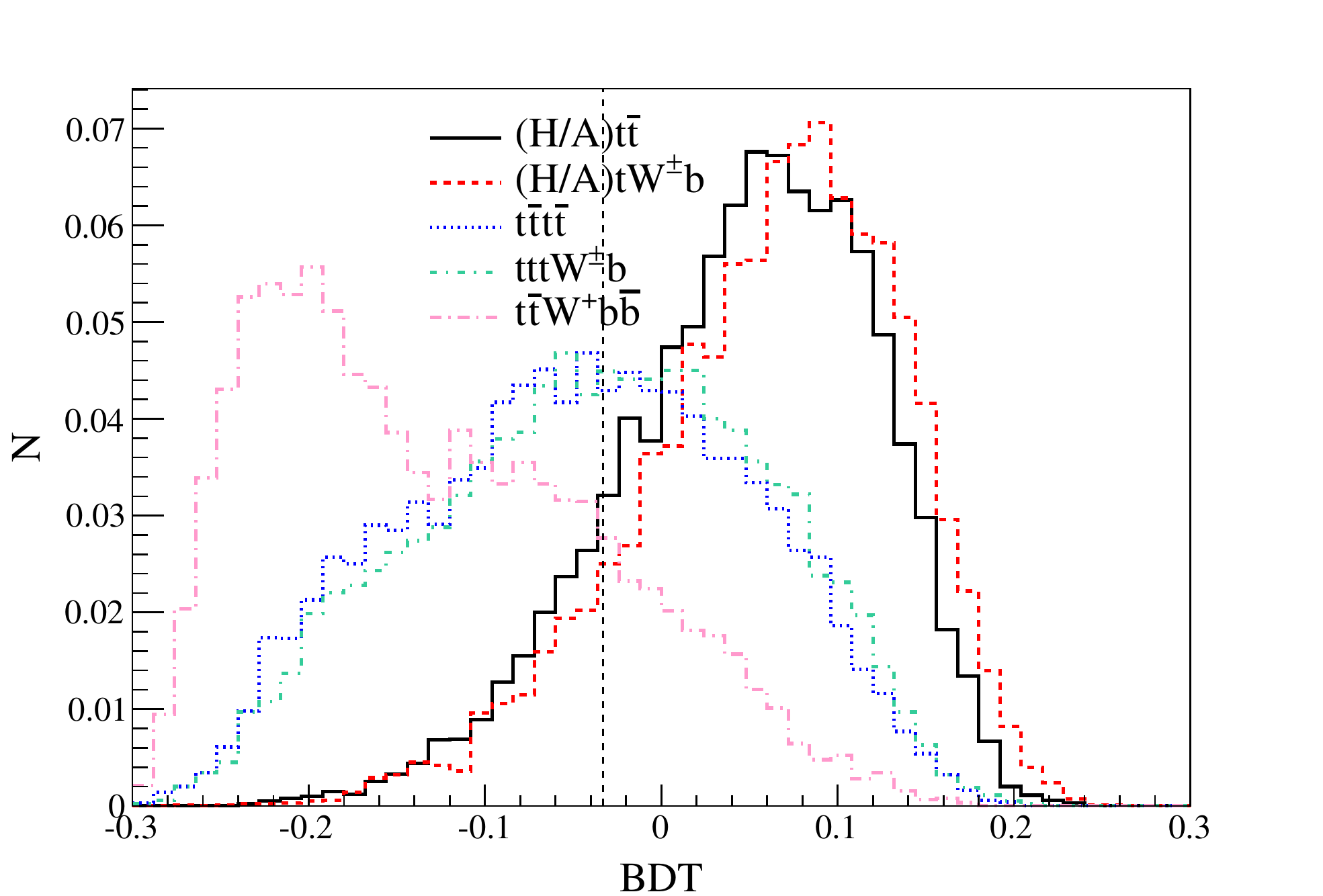}
\caption{%
Example of a BDT result for the \unit[14]{TeV} LHC and a Higgs mass of \unit[1]{TeV}.
While the Higgs produced in association with two (black) and one (red) top quark has large BDT values,
the intrinsic background $t\bar tt\bar t$ (blue), $tttW^{\pm}b$ (green) and $ttW^{+} bb$ (pink) has small BDT values.
The vertical line indicates the optimal cut, which maximizes the significance.
}
\label{fig:BDT result}
\end{figure}

We focus the BDT-based search strategy on the the main features of this signal, namely
\begin{itemize}
 \item Same sign di-lepton (SSDL)
 \item Heavy Higgs boson resonance
 \item Two additional forward/backward $b$-quarks
\end{itemize}
We apply a series of BDT-taggers designed to reconstruct the complete signal signature from its decay components following the strategy presented in~\cite{Hajer:2015gka}.
Our code is based on \software[3.1.3]{FastJet}~\cite{Cacciari:2011ma} and the \software[4.2.1]{TMVA} library~\cite{Hocker:2007ht} of the \software{ROOT} framework~\cite{Brun:1997pa} and is published as \software[0.2]{BoCA}~\cite{Boca}.
We tag bottom-like jets, based on their displaced vertices.
We exploit the fact that boosted top-like jets also show displacement but have additionally a larger jet mass.
Furthermore we require or veto hard leptons inside the jet radius for hadronic and leptonic top jets, respectively.
In the case of un-boosted top quarks we reconstruct the tops from their spatially separated decay products.
For the reconstruction of the heavy Higgs bosons we require two tagged top-like objects.
We do not reconstruct the tops of the quark pair accompanying the heavy Higgs.
Instead we require two jets with large $\Delta \eta$ and high $b$-likeliness.
Finally we assume that one of the tops originating from the heavy Higgs decay, decays leptonically; that another lepton is present; and that these leptons have the same charge.
One example of the final BDT result at \unit[14]{TeV} is presented in Figure~\ref{fig:BDT result}.

\section{Results} \label{sec:prospects}

Here we present the results of the cut-based and BDT analyses presented in the previous section. We note that in Section \ref{sec:constraints} we presented limits individually for scalar and pseudoscalar Higgs bosons in the context of a simplified model, as \unit[8]{TeV} data is insufficient to meaningfully constrain the parameter space of the MSSM. In what follows, we present limits for the sum of scalar and pseudoscalar Higgs bosons in the parameter space of $m_A$ and $\tan \beta$. We do so with an eye towards forecasting sensitivity to MSSM-like type~II 2HDM in which the heavy scalar and pseudoscalar Higgs bosons are nearly degenerate.

\subsection{Cut-based and BDT analyses at the \unit[14]{TeV} LHC} \label{sec:BDT-14}

\begin{figure}
\begin{subfigure}{.5\textwidth}
\includegraphics[width=\textwidth]{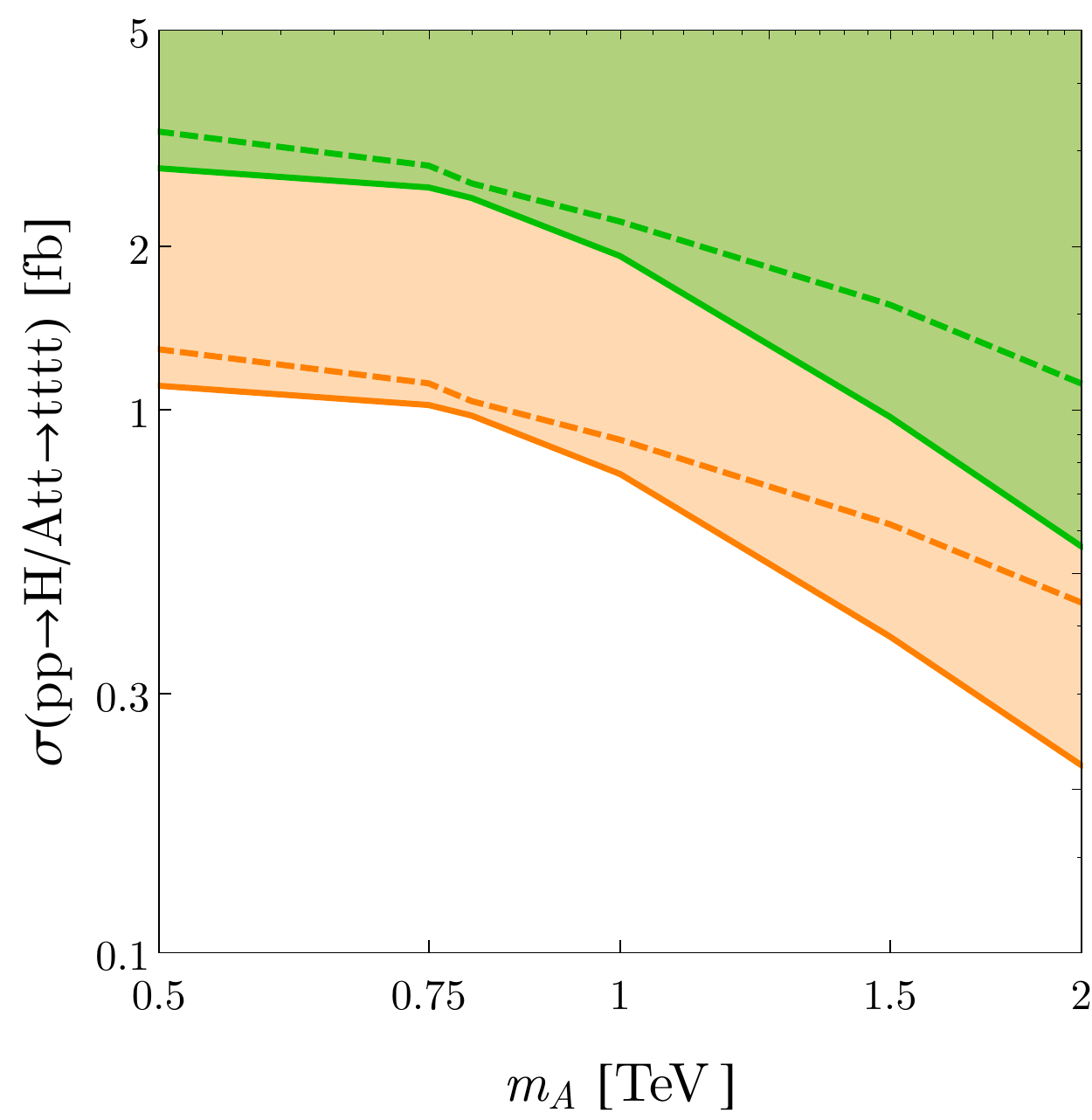}
\caption{$pp\to t\bar tH(A)\to t\bar tt\bar t$}
\label{fig:IndependentLimitLhc4t}
\end{subfigure}
\begin{subfigure}{.5\textwidth}
\includegraphics[width=\textwidth]{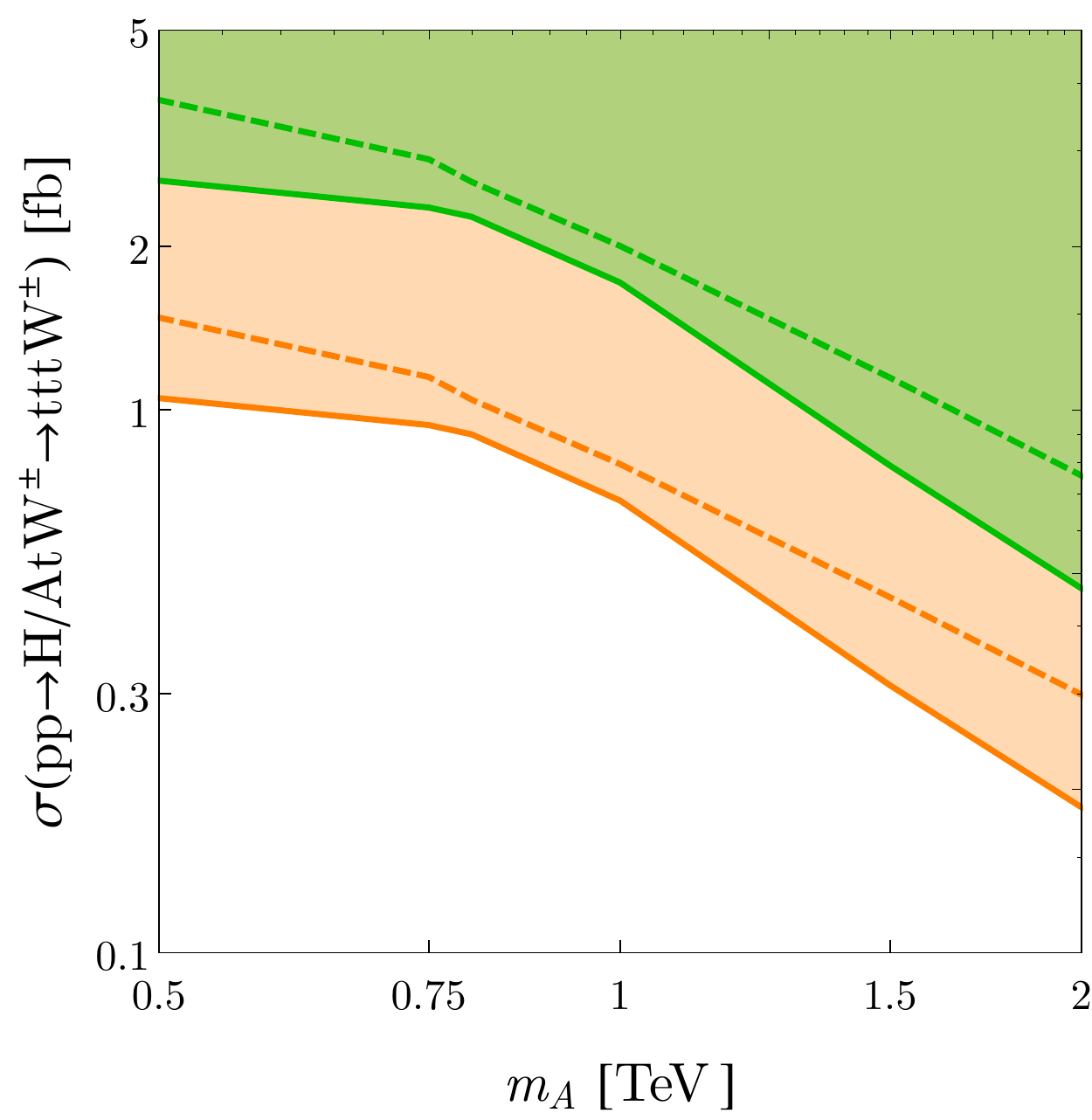}
\caption{$pp\to tWH(A)\to tW^\pm t\bar t$}
\label{fig:IndependentLimitLhc3t}
\end{subfigure}
\caption{%
(\subref{fig:IndependentLimitLhc4t}) Model independent exclusion (orange) and discovery (green) limits at the \unit[14]{TeV} LHC in the four-top channel. (\subref{fig:IndependentLimitLhc3t}) Exclusion (orange) and discovery (green) limits in the three-top channel.
The dashed limits are derived with the cut based analysis presented in Section~\ref{sec:cut-based} while the solid limits are derived with the BDT analysis presented in Section~\ref{sec:BDT}.
}
\label{fig:IndependentLimitLhc}
\end{figure}

\begin{figure}
\begin{subfigure}{.5\textwidth}
\includegraphics[width=\textwidth]{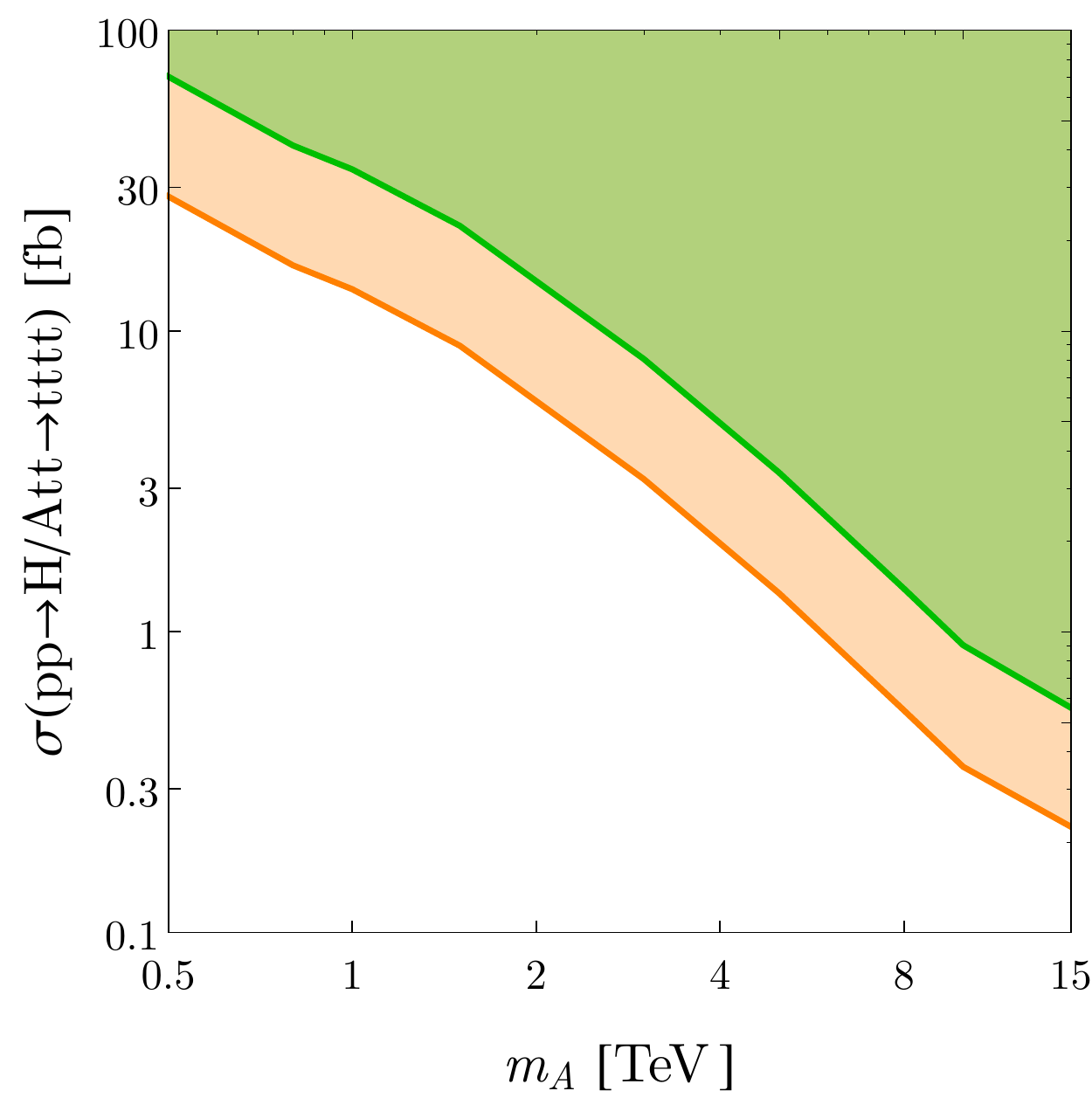}
\caption{$pp\to t\bar tH(A)\to t\bar tt\bar t$}
\label{fig:IndependentLimit4t}
\end{subfigure}
\begin{subfigure}{.5\textwidth}
\includegraphics[width=\textwidth]{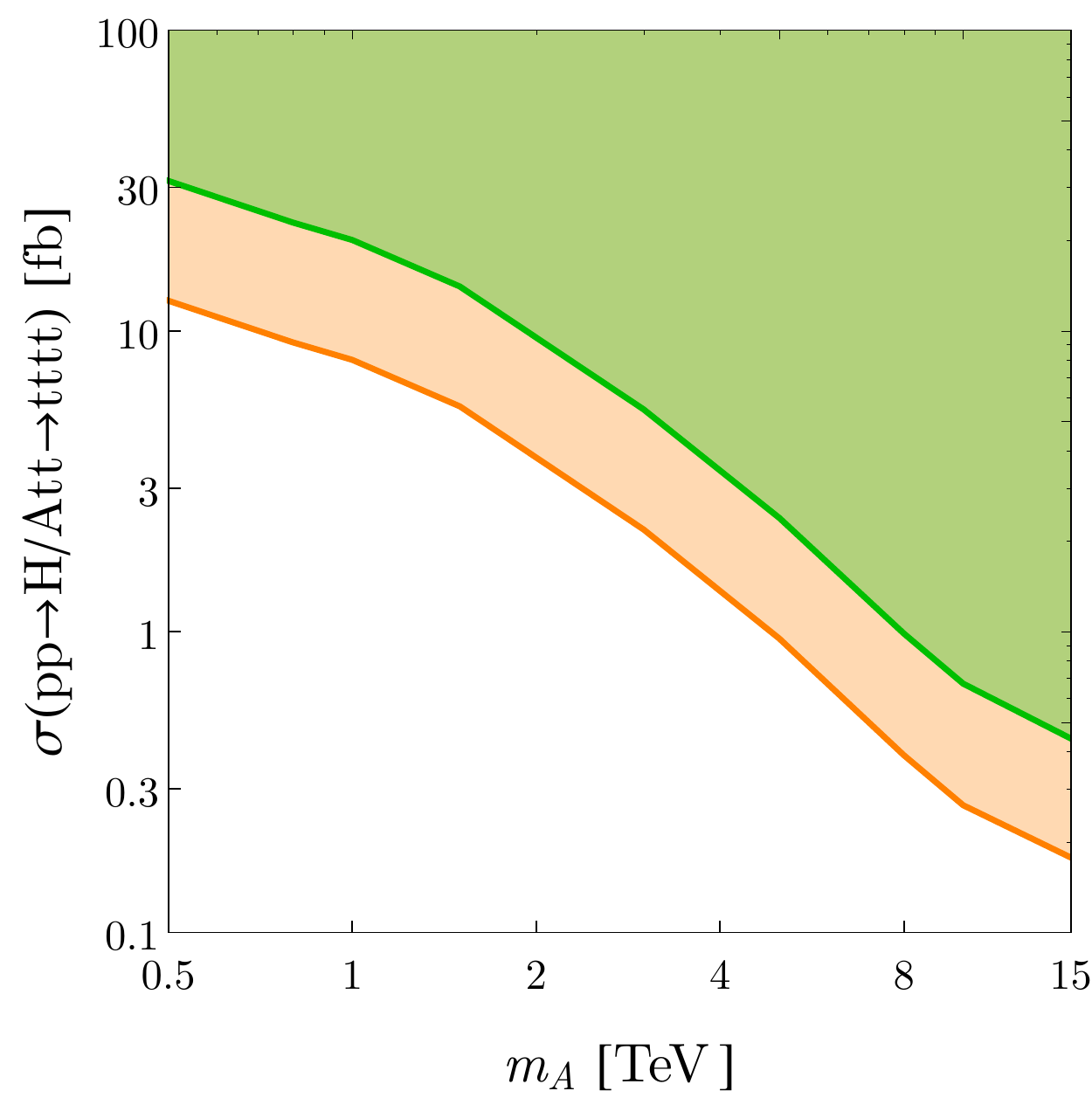}
\caption{$pp\to tW^\pm H(A)\to tW^\pm t\bar t$}
\label{fig:IndependentLimit3t}
\end{subfigure}
\caption{%
(\subref{fig:IndependentLimit4t}) Model independent limits for a future \unit[100]{TeV} collider in the $t\bar tt\bar t$ channel. (\subref{fig:IndependentLimit3t}) Limits for a future \unit[100]{TeV} collider in the $t\bar ttW^\pm$~ channel.
Note the dominance of the three-top channel over the four-top channel especially for low masses, due to the harder lepton originating from the $W^\pm$ decay compared to the top decay (cf.~Figure~\ref{fig:lepton pt}).
}
\label{fig:IndependentLimit-100}
\end{figure}

\begin{figure}
\begin{subfigure}{.41\textwidth}
\includegraphics[width=\textwidth]{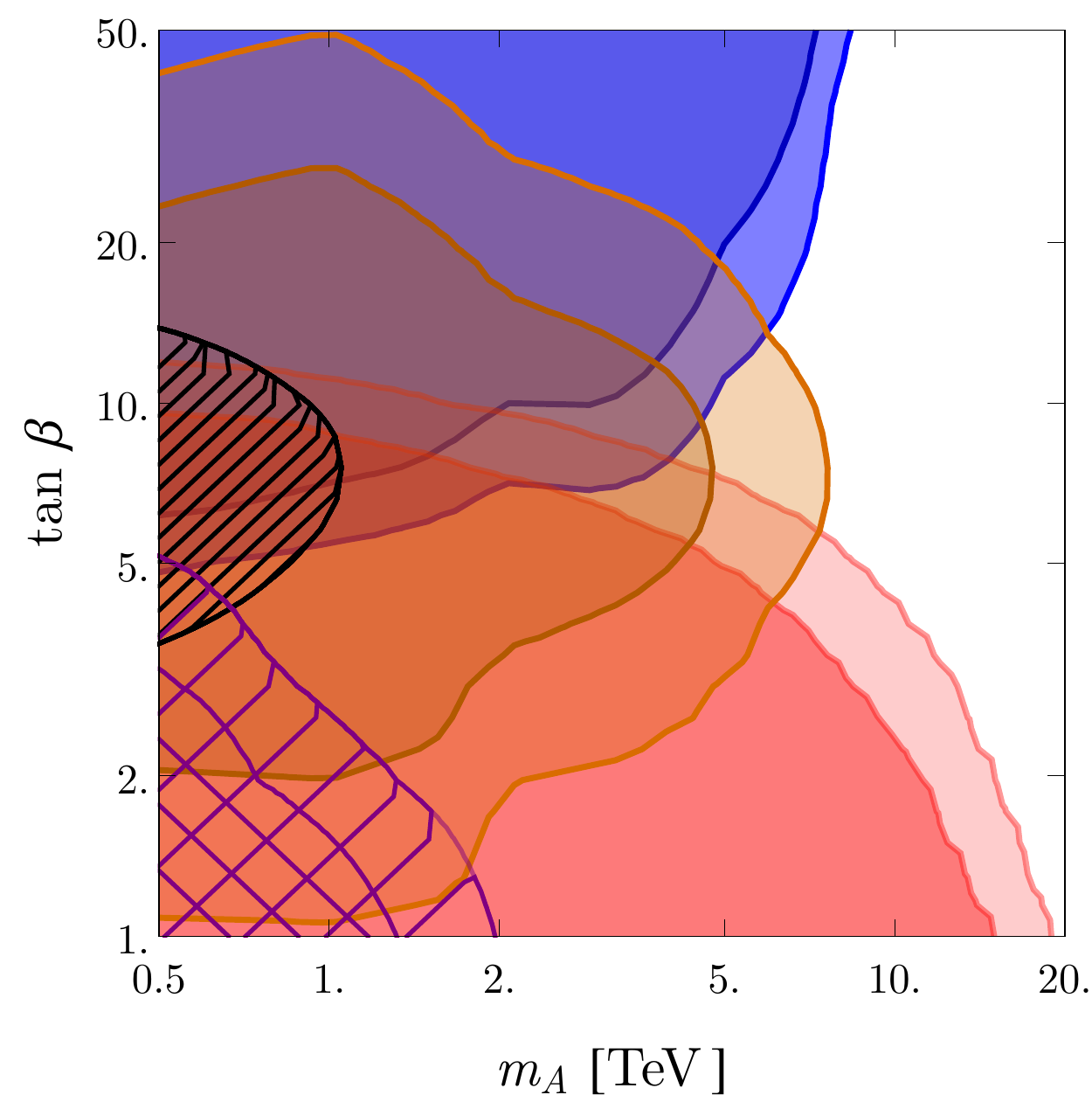}
\caption{Exclusion limits}
\label{fig:DependentExclusionLimit}
\end{subfigure}
\begin{subfigure}{.18\textwidth}
\includegraphics[width=\textwidth]{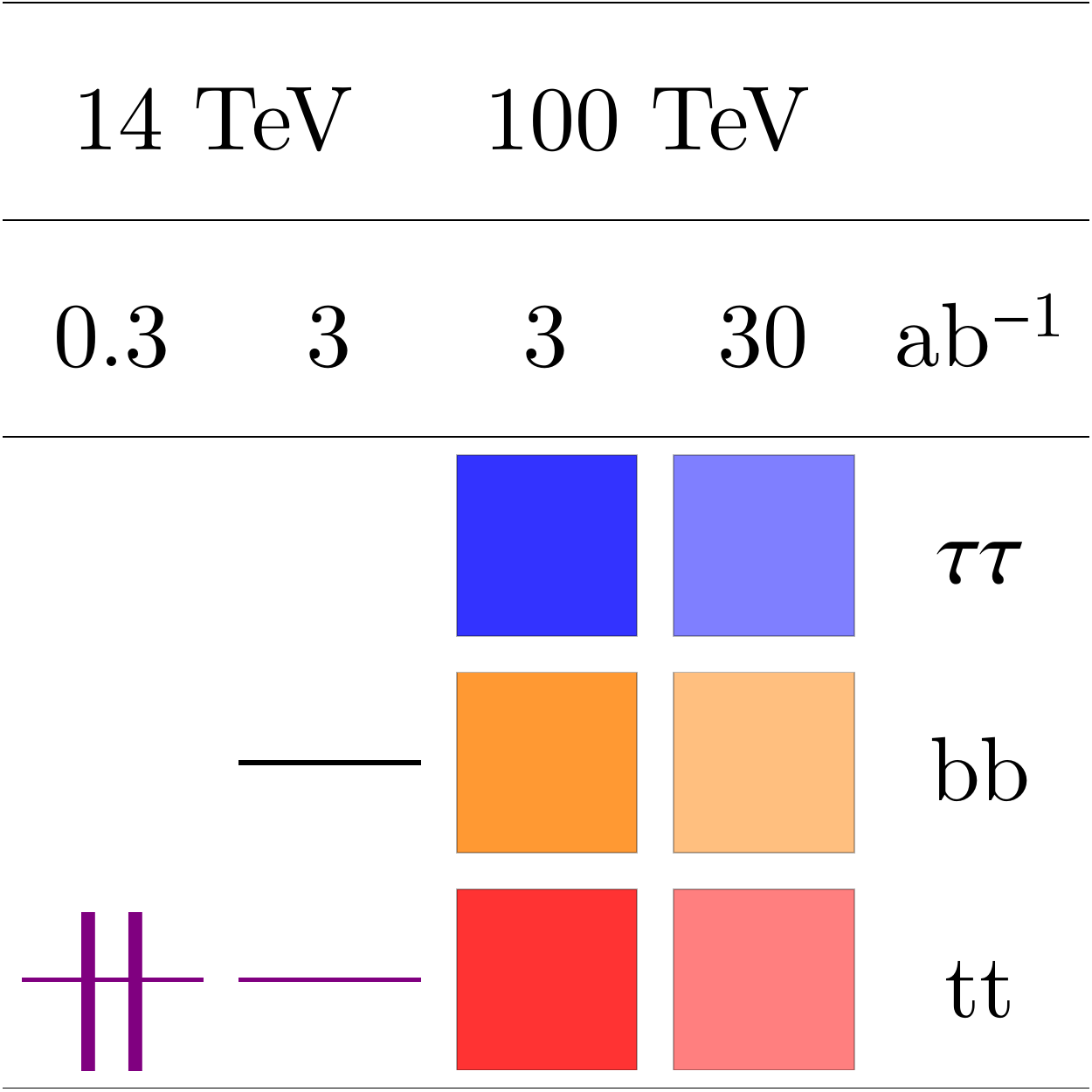}
\end{subfigure}
\begin{subfigure}{.41\textwidth}
\includegraphics[width=\textwidth]{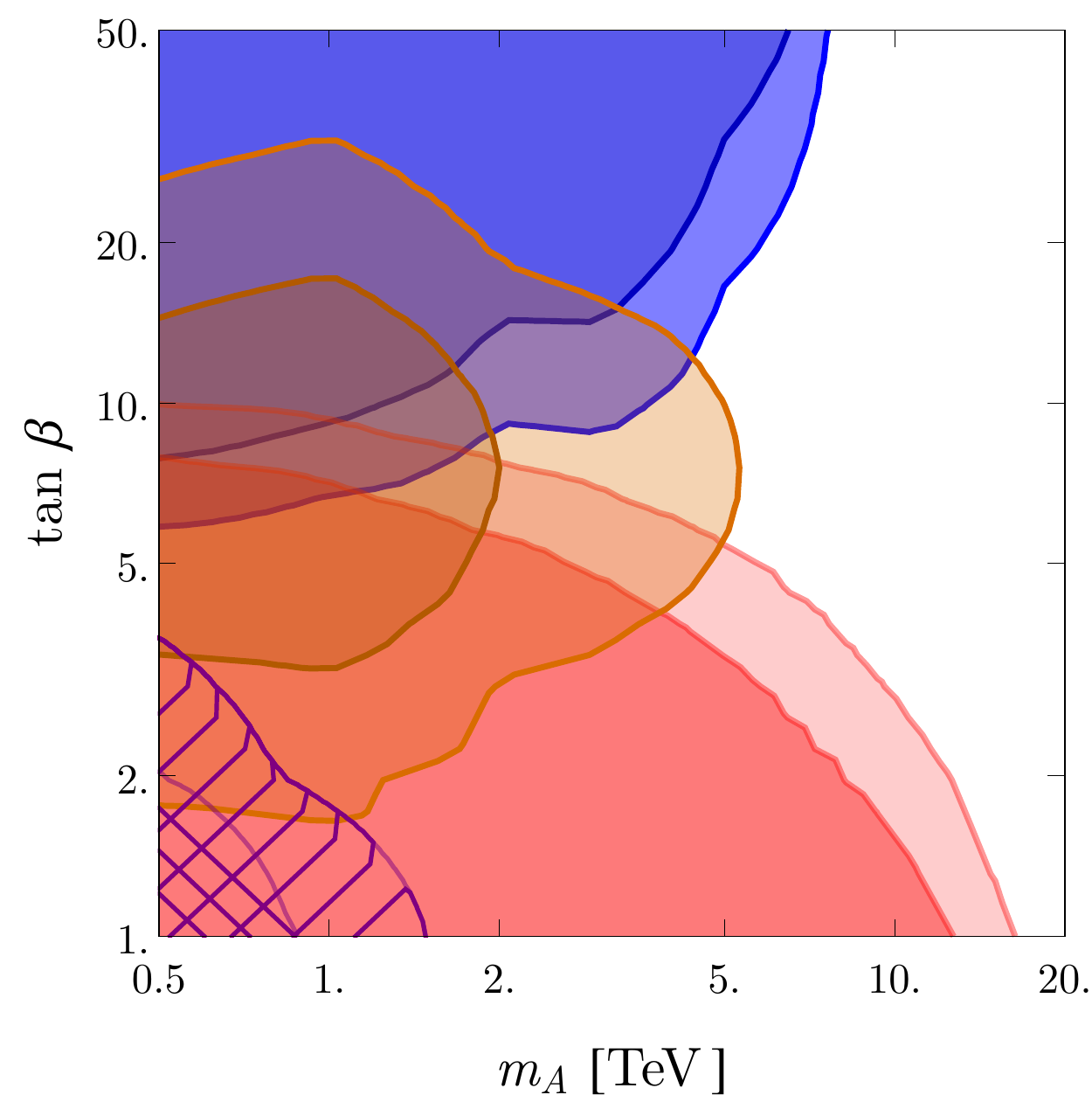}
\caption{Discovery limits}
\label{fig:DependentDiscoveryLimit}
\end{subfigure}
\caption{%
Model dependent exclusion~(\subref{fig:DependentExclusionLimit}) and discovery~(\subref{fig:DependentDiscoveryLimit}) limits for the \unit[14]{TeV} LHC (hatched in black and purple) and a \unit[100]{TeV} hadron collider (colored) derived with the BDT analysis presented in Section~\ref{sec:BDT}.
The smaller bound can be reached with $\unit[0.3 \text{ and } 3]{ab^{-1}}$ while the large bound can be reached with $\unit[3 \text{ and } 30]{ab^{-1}}$ at the LHC and a future $pp$-collider, respectively.
The low $\tan\beta$ region (red) is covered by the top associated heavy Higgs production with decays to top pairs.
While the contribution from the $H/A b\bar b$ vertex dominates the decays for large $\tan\beta$ we neglected its sub-leading contribution in the analysis covering small $\tan\beta$.
The intermediate $\tan\beta$ region (orange) is covered by the bottom associated heavy Higgs production with decays to a top pair.
The large $\tan\beta$ region (blue) is covered by the bottom associated heavy Higgs production with decays to $\tau$ lepton pairs.
The latter two analyses are discussed in~\cite{Hajer:2015gka}, and we revisit some aspects of the analysis in Appendix~\ref{sec:systematic error}.
}
\label{fig:DependentLimit}
\end{figure}

The model-independent results for cut- and BDT-based analyses at the \unit[14]{TeV} LHC are depicted separately for the three- and four-top channels in Figure~\ref{fig:IndependentLimitLhc}.
While the cut-based and BDT analyses have comparable sensitivity for lower masses, for larger masses the BDT-based strategy demonstrates its power as the reconstructed heavy Higgs resonance improves discrimination of signal from background.
Furthermore, we would like to point out that the BDT-based analysis has an advantage in suppressing the signal over background ratio, as discussed in Appendix~\ref{sec:systematic error}.
We present the combined model dependent results of the BDT analysis in Figure~\ref{fig:DependentLimit}; the shaded purple regions denote the reach of the \unit[14]{TeV} LHC with $\unit[300 \text{ and } 3000]{fb^{-1}}$.
As shown in Figure~\ref{fig:DependentExclusionLimit} the top associated heavy Higgs production can exclude the lower $\tan\beta$ region up to \unit[1 and 1.8]{TeV} for $\unit[300 \text{ and } 3000]{fb^{-1}}$, respectively.
The discovery reach of the LHC presented in Figure~\ref{fig:DependentDiscoveryLimit} extends to \unit[700 and 1100]{GeV} for the same luminosities.
As already demonstrated in~\cite{Hajer:2015gka} the bottom associated heavy Higgs production covers the intermediate $\tan\beta$ region up to \unit[1]{TeV} with $\unit[3000]{fb^{-1}}$.%
\footnote{The slight change in the shape of the limits derived from the bottom associated Higgs production with decays to a top pair compared to~\cite{Hajer:2015gka} is the result of a combination of improved background simulation and correspondingly optimized analysis.}

\subsection{BDT analysis at a \unit[100]{TeV} $pp$-Collider}

The model-independent sensitivity of a \unit[100]{TeV} collider is presented separately for the three- and four-top channels in Figure~\ref{fig:IndependentLimit-100}.
The model-dependent combination of three- and four-top channels is depicted alongside the \unit[14]{TeV} LHC reach in Figure~\ref{fig:DependentLimit}.
The top associated heavy Higgs production can exclude the lower $\tan\beta$ range up to \unit[15 and 18]{TeV} for $\unit[3 \text{ and } 30]{ab^{-1}}$, respectively.
The discovery reach extends to \unit[10 and 15]{TeV} for the same luminosities.
Of course, large uncertainties regarding detector properties, backgrounds, and BDT performance at \unit[100]{TeV} make these limits approximate.
The complementary bottom associated heavy Higgs production mode can be used to exclude the intermediate $\tan\beta$ region up to \unit[4 and 8]{GeV} for $\unit[3 \text{ and } 30]{ab^{-1}}$, respectively.
Finally the associated heavy Higgs production with two bottom quarks and decays to a $\tau$ lepton pair covers the large $\tan\beta$ range.
Together, these channels cover the whole $\tan\beta$ range up to $\sim \unit[10]{TeV}$.

Combining the dominance of the three-top channel over the four-top channel in Figure~\ref{fig:IndependentLimit-100} with the larger cross-section of the three-top channel compared to the four-top channel observed in Figure~\ref{fig:ProductionXsectiont},  the $H(A)W^\pm b$ channel provides the main contribution to the limits presented in Figure~\ref{fig:DependentLimit}.

\section{Summary and Outlook}\label{sec:summary}

Heavy Higgs bosons decaying predominantly into $t \bar t$ final states pose an exceptional challenge to searches at hadron colliders, particularly when $b \bar b$ associated production is negligible. This makes it difficult to probe a variety of motivated theories with heavy Higgs bosons decaying to $t \bar t$, including most notably the low $\tan\beta$ region of the MSSM Higgs sector.
In this work we have proposed probing this parameter space by searching for heavy Higgses produced in association with one and two top quarks. While these processes may be probed in a variety of final states, we have focused on searches involving same-sign dilepton pairs.
We have shown that existing LHC searches at \unit[8]{TeV} are sensitive to these channels, though current limits are too weak to meaningfully constrain the parameter space of heavy Higgs bosons.
However, the refinement of these cut-based searches at \unit[14]{TeV} will begin to meaningfully constrain the relevant parameter space. Further improvement with BDT-based analyses can potentially cover the lower part of the $m_A$-$\tan\beta$ plain up to \unit[700--1800]{GeV} and \unit[10--18]{TeV} at the LHC and a future hadron collider, respectively.
Together with the results of~\cite{Craig:2015jba, Hajer:2015gka} we have shown that the complete $\tan\beta$ range can be covered with associated heavy Higgs production, while only the upper part of this range can be probed up to large masses with resonantly-produced heavy Higgs bosons.
Although the sensitivity reach in this article was displayed in type II 2HDM, the proposed strategies can be used to probe the low $\tan \beta$ region of other 2HDM as well.
The sensitivity projection is straightforward.

\subsection*{Acknowledgment}

J. Hajer is supported by the Collaborative Research Fund (CRF) HUKST4/CRF/13G.
Y.-Y. Li is supported by the Hong Kong PhD Fellowship Scheme (HKPFS).
T. Liu is supported by the General Research Fund (GRF) under Grant No. 16304315 and 16312716.
Both the HKPFS and the CRF, GRF grants are issued by the Research Grants Council of Hong Kong S.A.R..
N. Craig is supported by the U.S. DOE under contract No. DE-SC0014129. H. Zhang is supported by the U.S. DOE under contract No. DE-SC0011702.

\appendix

\section{Acceptances of the \software{BoCA} SM Taggers}
\label{sec:boca tagger}

In order to test the \software{BoCA} SM taggers, we have generated event files containing particle pairs generated via $t$- and $s$-channel contributions.
During the analysis we have required exactly one particle and one jet to fall inside the required transverse momentum window.

\begin{figure}
\begin{subfigure}{.5\textwidth}
\includegraphics[width=\textwidth]{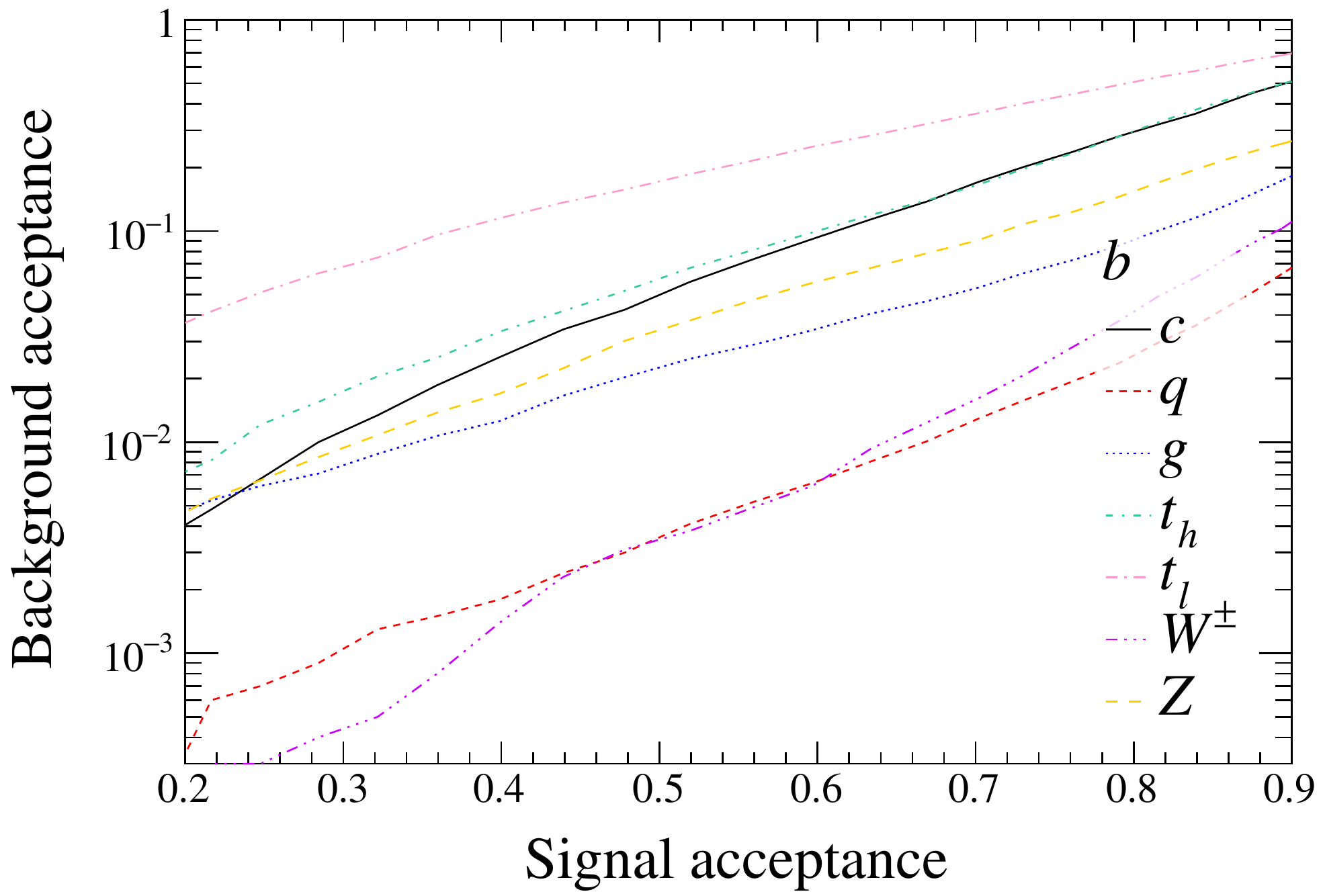}
\caption{Acceptances}
\label{fig:bottom acceptances}
\end{subfigure}
\begin{subfigure}{.5\textwidth}
\includegraphics[width=\textwidth]{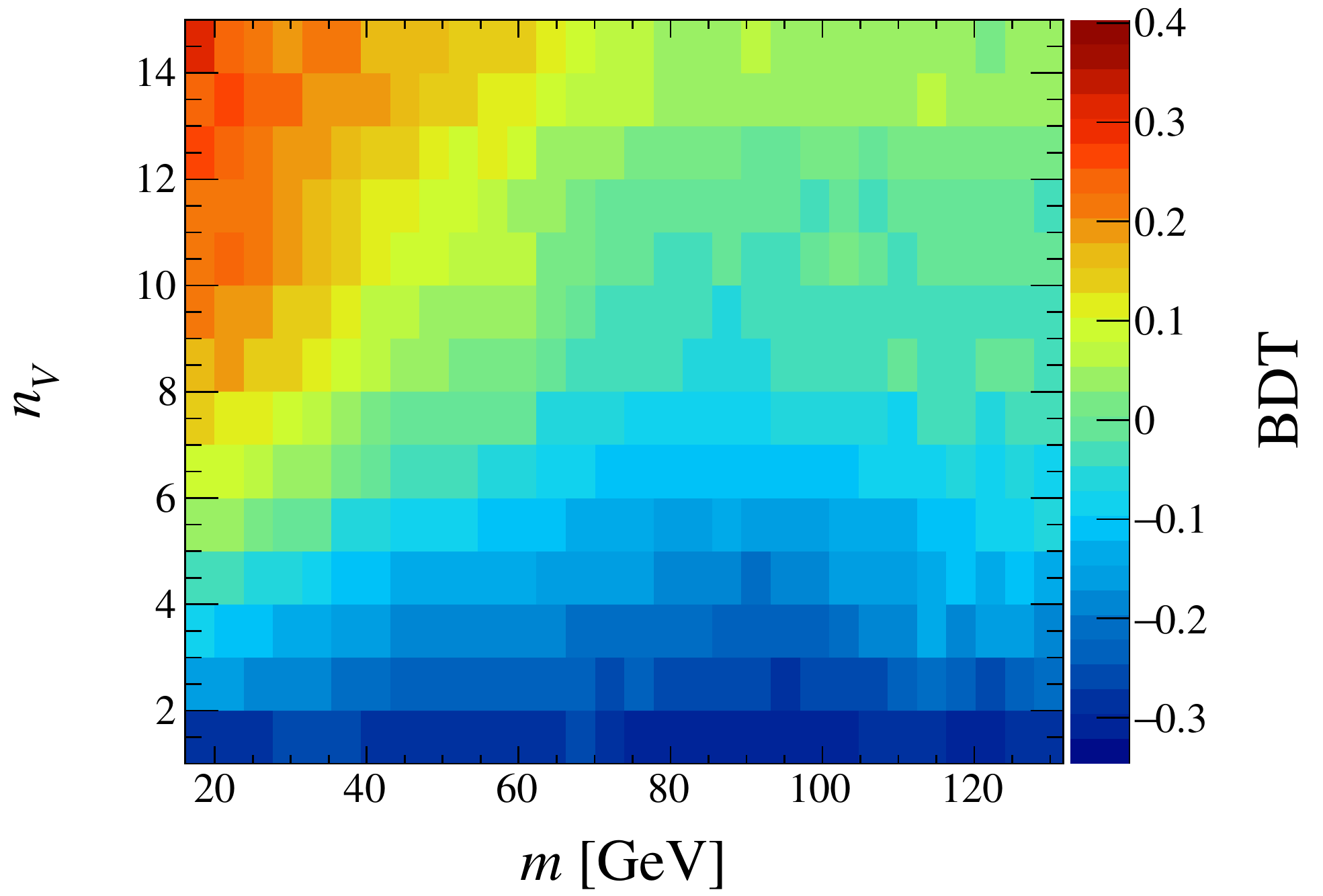}
\caption{BDT over mass and multiplicity.}
\label{fig:bottom bdt}
\end{subfigure}
\caption{%
\software{BoCA} bottom tagger. (\subref{fig:bottom acceptances}) Acceptances for jets with a transverse momentum of $\unit[500]{GeV} < p_T < \unit[600]{GeV}$ at the \unit[14]{TeV} LHC.
(\subref{fig:bottom bdt}) Distribution of averaged BDT values after projection of the parameter space onto the plane spanned by the jet mass ($m$) and the track multiplicity ($n_V$).
Low BDT values (blue) are background-like and high BDT values (red) are signal-like.
We require the final state particles to have transverse momenta larger than \unit[500]{GeV} to make sure that the induced two
jets are well-separated.}
\label{fig:bottom boca}
\end{figure}

The \software{BoCA} bottom tagger exploits the longevity of $B$-mesons produced in the bottom decay and uses the radial displacement of jet tracks in order to discriminate against other jets.
The track positions provided by \software{Delphes} are smeared and serve as a starting point for the calculation of displacement observables, such as the track multiplicity, average displacement of tracks and the invariant mass of the displaced tracks.
We do not attempt to reconstruct secondary jet vertices.
The signal and background acceptances for jets with $\unit[500]{GeV} < p_T < \unit[600]{GeV}$ as well as a representation of the BDT response are presented in Figure~\ref{fig:bottom boca}.
Using a BDT-cut of 0.05 leads to a $b$-tagging rate of \unit[71]{\%} with a $c$-jet fake rate of \unit[18]{\%}, while the misidentification rate for light quarks and gluon jets is \unit[1.4]{\%} and \unit[6.4]{\%}, respectively.

\begin{figure}
\begin{subfigure}{.5\textwidth}
\includegraphics[width=\textwidth]{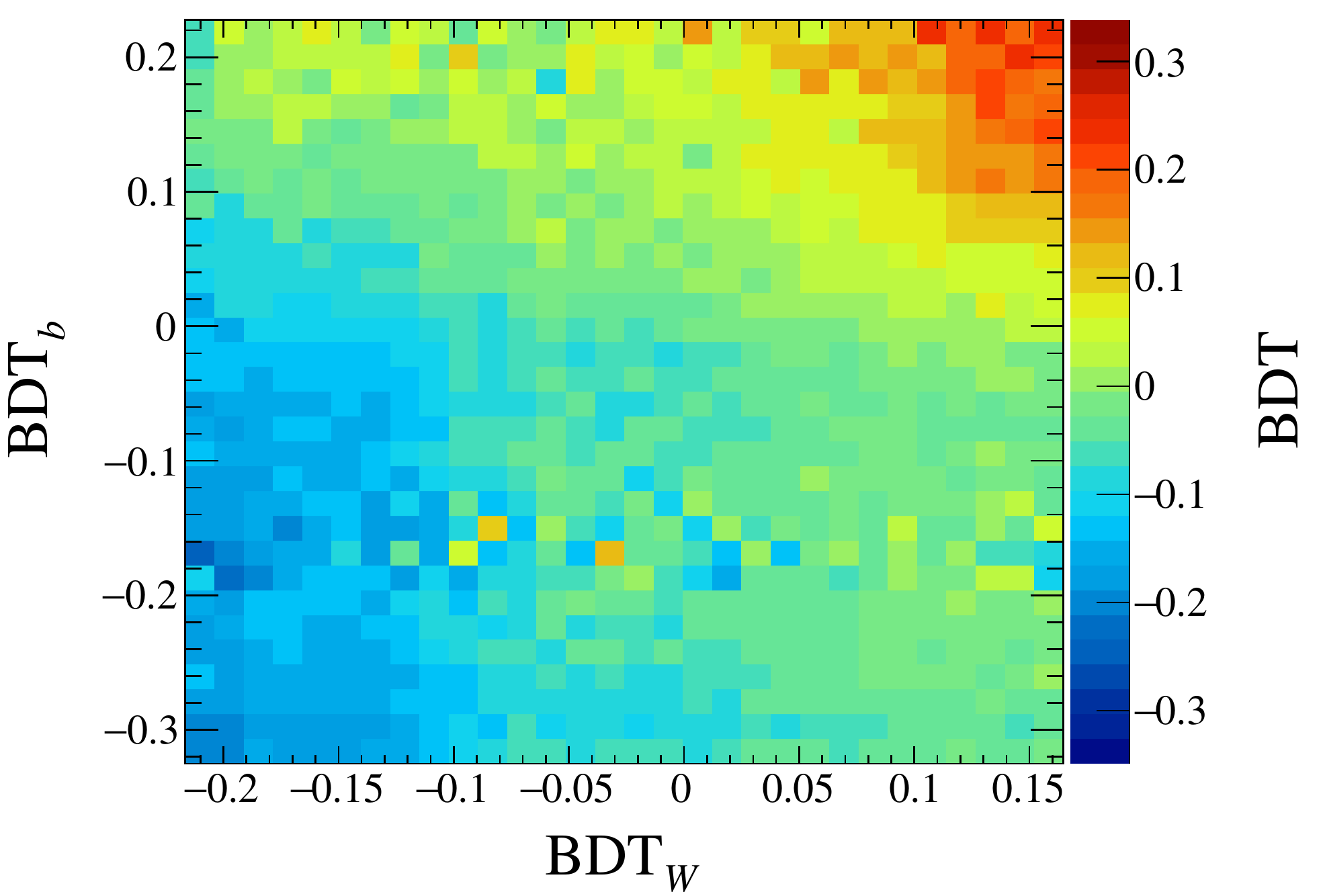}
\caption{$W$ and $b$ BDT}
\label{fig:three BDT}
\end{subfigure}
\begin{subfigure}{.5\textwidth}
\includegraphics[width=\textwidth]{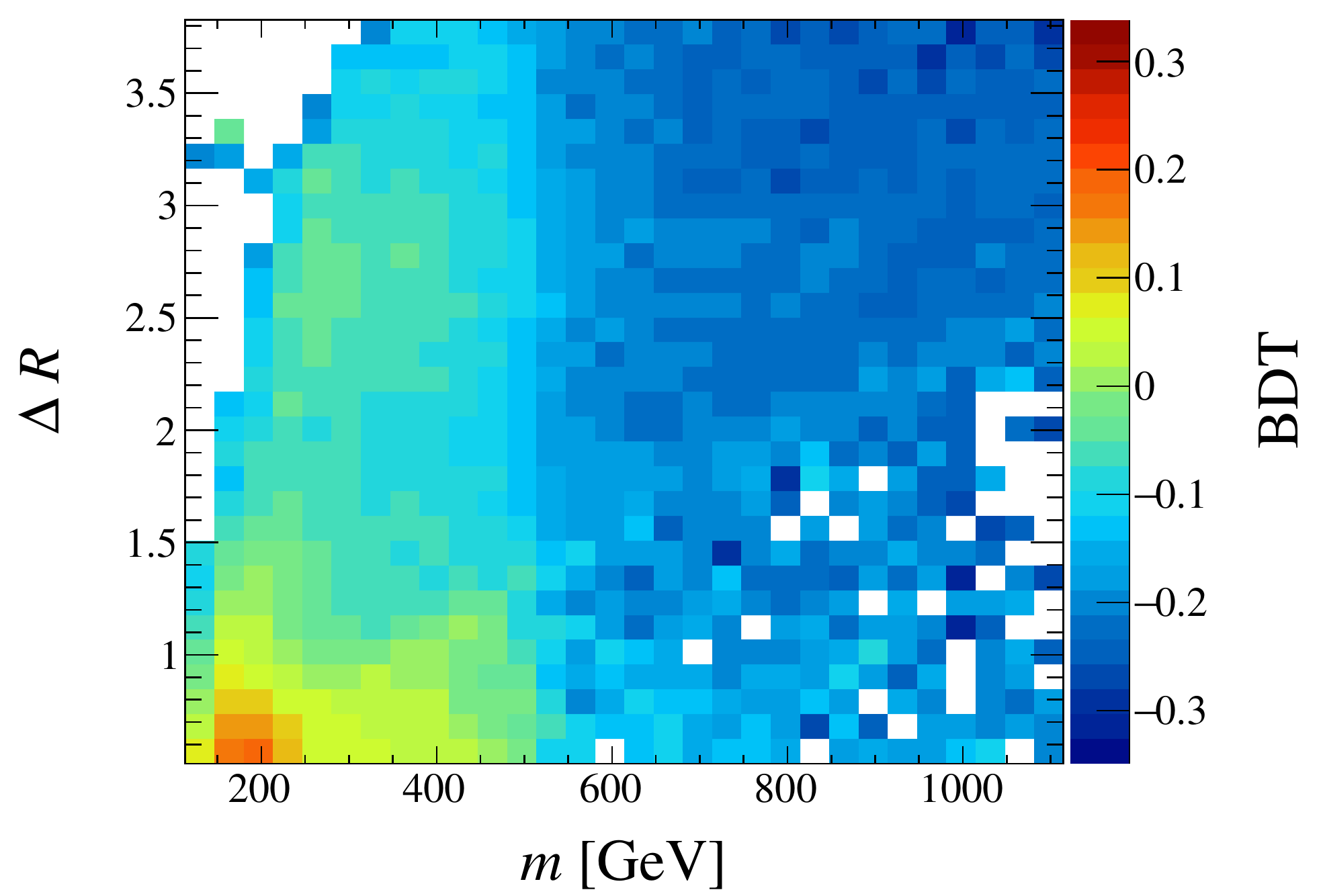}
\caption{Top mass and opening angle}
\label{fig:top mass opening angle}
\end{subfigure}
\caption{%
Examples of the top tagger BDT response. (\subref{fig:three BDT}) Averaged BDT response after projection of the parameter space onto the plane spanned by the $W$ BDT and the $b$ BDT. (\subref{fig:top mass opening angle}) Averaged BDT response after projection onto the plane spanned by the top mass ($m$) and the opening angle ($\Delta R$).
Low BDT values (blue) are background-like and high BDT values (red) are signal-like.
}
\label{fig:top examples}
\end{figure}

\begin{figure}
\begin{subfigure}{.5\textwidth}
\includegraphics[width=\textwidth]{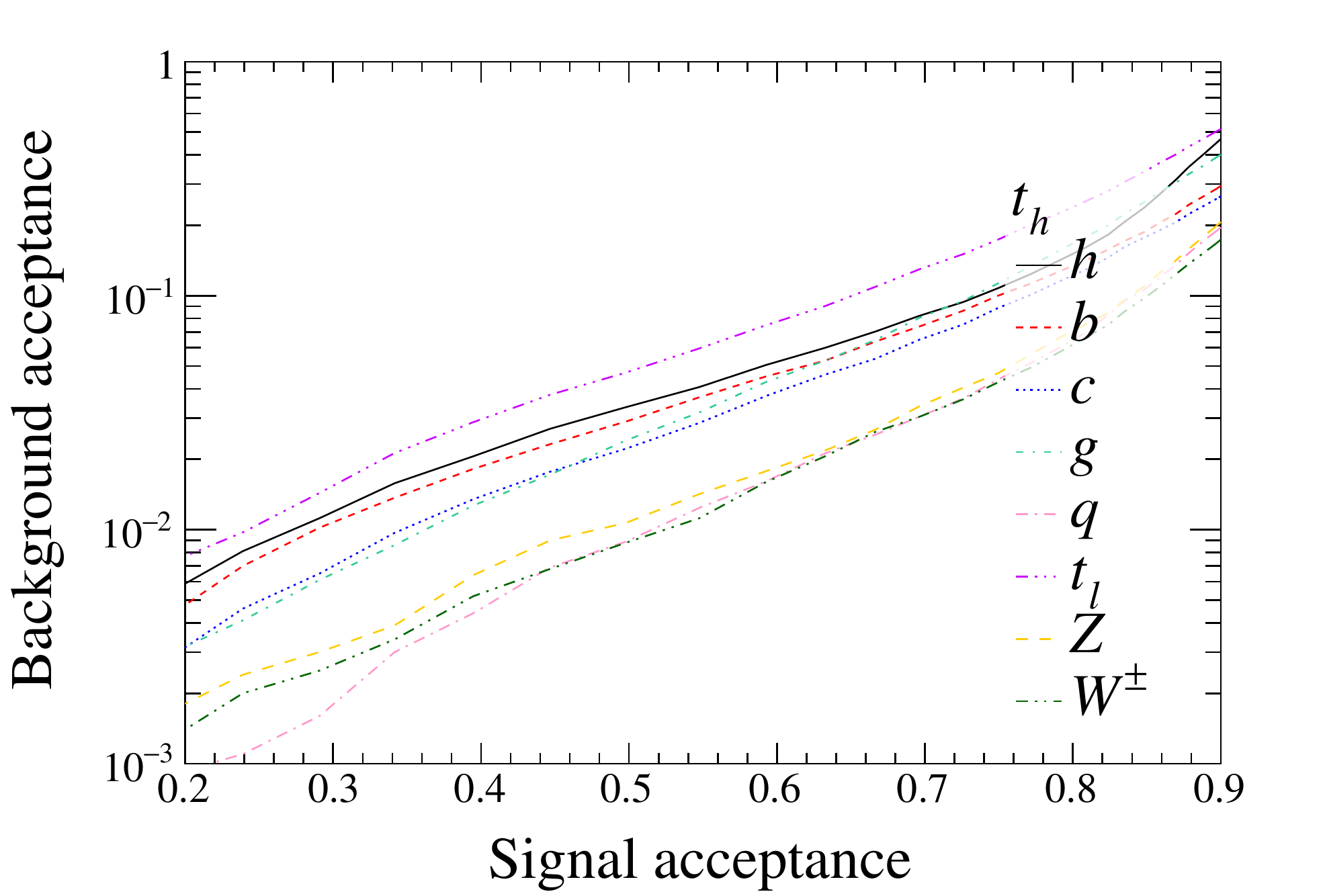}
\caption{\software{BoCA} top tagger.}
\label{fig:boca top tagger}
\end{subfigure}
\begin{subfigure}{.5\textwidth}
\includegraphics[width=\textwidth]{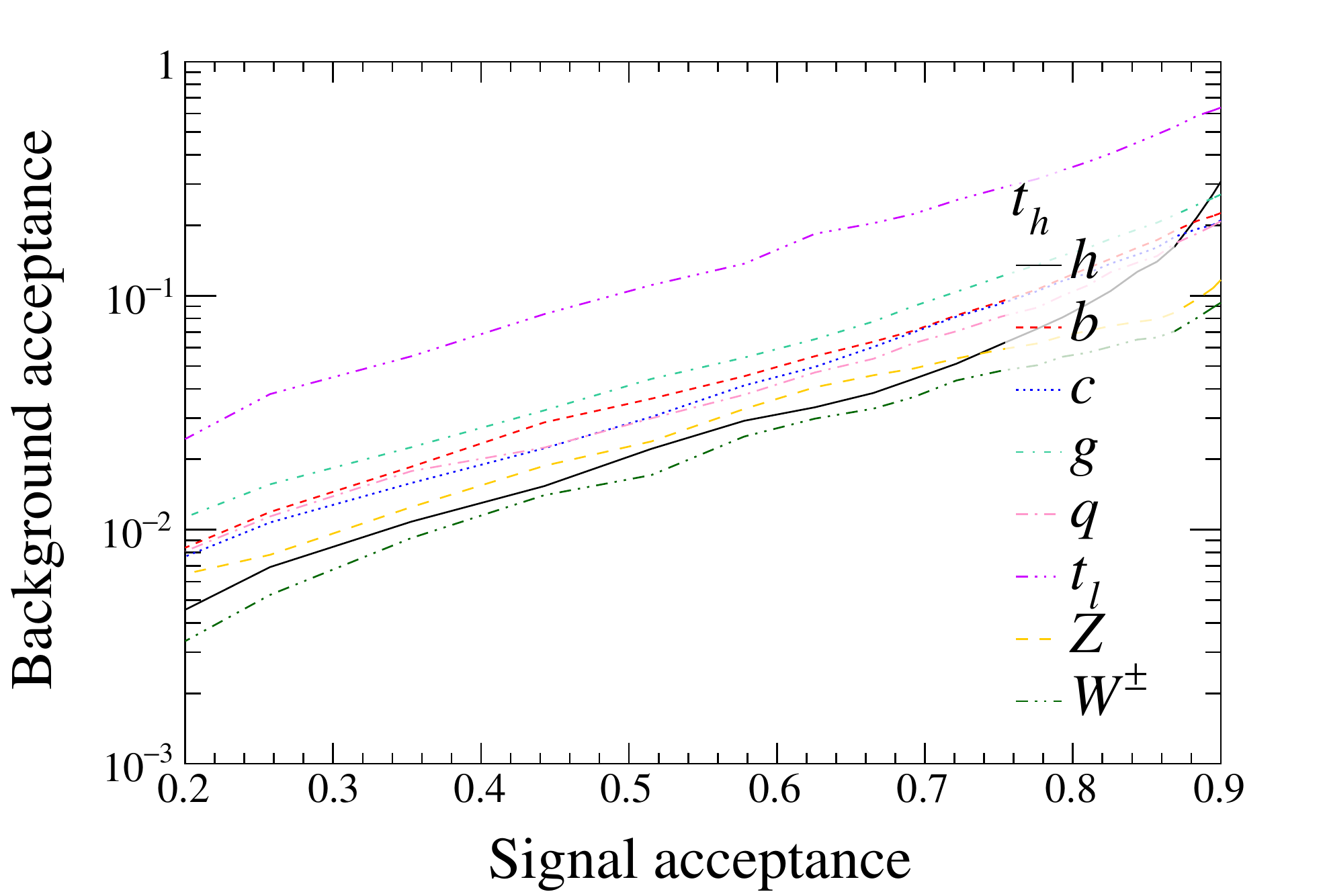}
\caption{\software{HEPTopTagger}.}
\label{fig:hep top tagger}
\end{subfigure}
\caption{%
Top tagger acceptances for top jets with an transverse momentum between $\unit[750]{GeV} < p_T < \unit[1]{TeV}$ at the \unit[14]{TeV} LHC.
(\subref{fig:boca top tagger}) Signal acceptance calculated using the \software{BoCA} top tagger with \software{Delphes} jets with a jet cone size of 0.5. (\subref{fig:hep top tagger}) Signal acceptance  calculated using the \software{HEPTopTagger} without a prior on the top mass using optimally clustered jets with a jet cone size of 0.462.
}
\label{fig:top tagger comparison}
\end{figure}

The \software{BoCA} tagger for hadronically decaying tops is based on the bottom tagger and a $W$-tagger.
Additionally, it takes the kinematic variables between these two particles into account, such as the rapidity difference and the opening angle $\Delta R$.
Two representations of the top tagger BDT response are shown in Figure~\ref{fig:top examples}.
In order to compare the performance of the \software{BoCA} top tagger to existing top taggers, we have incorporated the \software{HEPTopTagger}~\cite{Plehn:2010st} into the \software{BoCA} code.%
\footnote{During finalization of this article the code for the second version of the \software{HEPTopTagger}~\cite{Kasieczka:2015jma} has been published.
We leave the comparison of the \software{BoCA} top tagger to this new version for a future publication.
}
While the \software{HEPTopTagger} is optimized for top (fat-)jet tagging the \software{BoCA} top-tagger also combines multiple jets in order to reconstruct less boosted top jets.
For the purposes of comparison, we require exactly one top particle and one reconstructed top to have a transverse momentum between $\unit[750]{GeV} < p_T < \unit[1]{TeV}$.
In order to create a range of signal acceptances with the \software{HEPTopTagger} we initially do not cut on the top mass but use the freedom in this parameter to tune the signal acceptance.
The receiver operating characteristic (ROC) curves for the \software{BoCA} top tagger and the \software{HEPTopTagger} are presented in Figure~\ref{fig:top tagger comparison}.
The direct comparison in this energy range shows that the \software{BoCA} top tagger suppresses $b$ and $c$ jets as well as the \software{HEPTopTagger}, while it performs slightly better for gluon and $W$-boson jets.
For light quarks, $Z$- and Higgs-boson jets it performs roughly twice as well as the \software{HEPTopTagger}.

\section{Comparison between cut- and BDT-based approaches in $pp\to Hbb$}
\label{sec:systematic error}

\begin{figure}
\centering
\includegraphics[width=.9\textwidth]{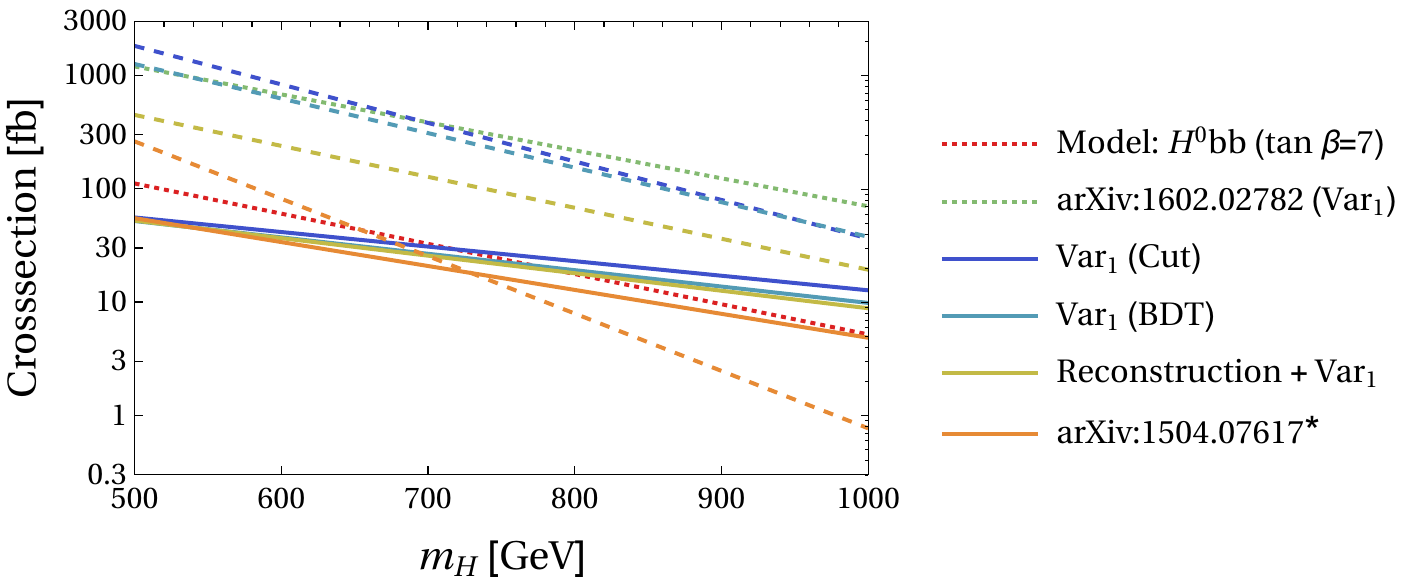}
\caption{%
Comparison of model independent exclusion cross section requiring a significance of $Z(b+s|b) \geq 2$ (solid) and a background suppression of $\frac sb \geq \unit[1]{\%}$ (dashed).
Applying the set of seven basic variables ($\text{Var}_1 = \{p_T(j_1)$, $p_T(j_2)$, $p_T(j_3)$, $p_T(j_4)$, $p_T(l_1)$, $H_T$, $\#b\}$) used in Reference~\cite{Gori:2016zto} (dotted green) leads in our analysis to comparable results with both cut- and BDT based approaches (dashed blue).
If we reconstruct the heavy Higgs boson and apply $\text{Var}_1$ (brown)
we are able to significantly increase the background suppression.
Using the technique described in~\cite{Hajer:2015gka} with the lepton isolation used in this publication (orange) leads once more to a notable improvement.
In~\cite{Hajer:2015gka} we have only considered hard leptons, in this analysis we use hard leptons as well as isolated soft leptons.
In the legend we use ``*'' to denote this difference.
}
\label{fig:model independent}
\end{figure}

In~\cite{Hajer:2015gka} we have analyzed the process $pp\to Hbb$ against the inclusive $t\bar t$ background.
The authors of the recent publication~\cite{Gori:2016zto} have attempted to exclude the same signal using a simple cut-based approach.
Their conclusion is that the small signal over background ratio makes it impossible to exclude relevant models even when $\frac{s}{\sqrt b}$ is appreciable.%
\footnote{%
The authors of~\cite{Gori:2016zto} point out that they found a $b$-quark distribution which differs from the one presented in~\cite{Hajer:2015gka}.
We have traced this difference back to deviations in the plot parameters. While we plot the $b$-quarks sorted by rapidity, the authors of~\cite{Gori:2016zto} plot the $b$-quarks sorted by their transverse momentum. We thank the authors of~\cite{Gori:2016zto} for clarifying their analysis.
}
Hence we would like to explicitly demonstrate that a BDT-based approach improves the background rejection ($\frac sb$) to the level demanded in~\cite{Gori:2016zto}.
In order to ensure comparability we use Figure~7 of~\cite{Gori:2016zto} as a starting point for our discussion.
The cross section of the model used in~\cite{Gori:2016zto} as well as the best exclusion limit derived in this paper are plotted as dotted lines in Figure~\ref{fig:model independent}.
The cut-based analysis in~\cite{Gori:2016zto} makes use of the transverse momenta of the four leading jets and the leading lepton, the scalar $H_T$, and the bottom number.
We call this set of variables $\text{Var}_1$ and have applied it in both a cut- and a BDT-based analysis.
Which leads to results comparable to those derived in~\cite{Gori:2016zto}.
In~\cite{Hajer:2015gka} we have reconstructed the complete event signature and used a large set of variables.
This set of variables contains the BDT of all reconstructed objects, the ratios between their kinematic observables, as well as the pull~\cite{Gallicchio:2010sw} between these objects.
Here we have only reconstructed the heavy Higgs boson and have applied $\text{Var}_1$
to suppress the background.
For low masses this improves the significance marginally, but the ratio between signal and background is increased significantly.
Finally, using the complete technique of~\cite{Hajer:2015gka} leads to a background suppression of $\order{\unit[0.2\text{--}6]{\%}}$ on the boundary of the exclusion region for the HL-LHC.
For a future \unit[100]{TeV} collider the background suppression on the boundary of the exclusion region varies in the range \unit[1--8]{TeV} between $\order{\unit[0.2\text{--}15]{\%}}$.

\bibliographystyle{JHEP}
\bibliography{references}

\end{document}